%% file: main_resubmitted_forastroph_nolinenumbers.tex
\documentclass[10pt,twocolumn,tighten]{aastex63}

\usepackage{amssymb,amsmath,times}
\usepackage{natbib}
\usepackage{graphicx}
\usepackage{graphbox}  
\usepackage{url}
\usepackage{color,xcolor}
\usepackage{textcomp}
\usepackage{rotating}
\usepackage[english]{babel}
\usepackage{breakurl}
\usepackage{graphicx,amssymb,amsmath,times,subfigure}
\usepackage{etoolbox}

\makeatletter

\usepackage{wasysym}

\begin{document}

\newcommand{\fermi}{\emph{Fermi}~}
\newcommand{\fermilat}{\emph{Fermi}-LAT~}

\newcommand{\vdag}{(v)^\dagger}
\newcommand{\aastex}{AAS\TeX}
\newcommand{\latex}{La\TeX}

\newcommand{\ltsima}{$\; \buildrel < \over \sim \;$}
\newcommand{\simlt}{\lower.5ex\hbox{\ltsima}} 
\newcommand{\gtsima}{$\; \buildrel > \over \sim \;$}
\newcommand{\simgt}{\lower.5ex\hbox{\gtsima}} 
\newcommand{\arcsecs}{\hbox{$^{\prime\prime}$}}
\newcommand{\degree}{\hbox{$^\circ$}}
\newcommand{\phflux}{photons cm$^{-2}$ s$^{-1}$}
\newcommand{\cgsflux}{erg s$^{-1}$ cm$^{-2}$}
\newcommand{\cgslum}{erg s$^{-1}$}
\newcommand{\gray}{$\gamma$-ray}
\newcommand{\grays}{$\gamma$-rays}
\newcommand{\dt}{\hbox{$\Delta$$t$}}
\newcommand{\dtr}{\hbox{$\Delta$$t_{\rm r}$}}
\newcommand{\dtg}{\hbox{$\Delta$$t_{\gamma}$}}
\newcommand{\Ms}{$M_\odot$}
\newcommand{\Ls}{$L_\odot$}
\newcommand{\Rs}{$R_\odot$}
\newcommand{\pzero}{$\pi^{\rm 0}$}
\newcommand{\Angst}{$\buildrel _{\circ} \over {\mathrm{A}}$}
\newcommand{\rninefive}{$r_{\rm 95}$}

\newcommand{\pg}{PG$~$1553+113}
\newcommand{\rxte}{\emph{RXTE}~}
\newcommand{\swift}{\emph{Swift}~}

\received{\today}
\revised{\today}
\accepted{\today}

\submitjournal{The Astrophysical Journal}


\shorttitle{Periodic Gamma-ray Modulation of the blazar PG 1553+113 Confirmed}
\shortauthors{(The \fermilat Collaboration) Abdollahi et al.}

\title{Periodic Gamma-ray Modulation of the blazar PG 1553+113\\ Confirmed by \textit{Fermi}-LAT and Multi-wavelength Observations}

\correspondingauthor{S. Ciprini, S. Cutini, S. Larsson, P. Cristarella Orestano}


\input{authorsOKfinal_4july2024.tex}

\begin{abstract}
A 2.1 year periodic oscillation of the $\gamma$-ray flux from the blazar PG 1553+113 has previously been tentatively identified in $\sim 7$ years of data from the \fermi Large Area Telescope. After 15 years of \fermi sky-survey observations, doubling the total time range, we report  $>7$-cycle $\gamma$-ray modulation with an estimated significance of 4 $\sigma$ against stochastic red noise. Independent determinations of oscillation period and phase in the earlier and the new data are in close agreement (chance probability $< 0.01$). Pulse timing over the full light curve is also consistent with a coherent periodicity. Multi-wavelength new data from {\it Swift} XRT, BAT and UVOT, and from KAIT, CSS, ASAS-SN, and OVRO ground-based observatories as well as archival {\it RXTE}-All Sky Monitor data, published optical data of Tuorla, and optical historical Harward plates data are included in our work. Optical and radio light curves show clear correlations with the $\gamma$-ray modulation, possibly with a non-constant time lag for the radio flux.
We interpret the gamma-ray periodicity as possibly arising from a pulsational accretion flow in a sub-parsec binary supermassive black hole system of elevated mass ratio, with orbital modulation of the supplied material and energy in the jet. Other astrophysical scenarios introduced, include  instabilities, disk and jet precession, rotation or nutation and perturbations by massive stars or intermediate mass black holes in polar orbit.

\end{abstract}


  \section{Introduction} \label{sect:intro}

Blazars are active galactic nuclei (AGNs), harboring accreting supermassive black holes (SMBHs) with powerful relativistic jets almost aligned with our line of sight \citep{urry}.
One of the more remarkable such objects is \pg\, a high-energy-peaked BL Lac object which is observed over the full electromagnetic spectrum, from radio to very high energy (VHE, $>$100 GeV) $\gamma$ rays. \pg\ was discovered as an optical and X-ray source \citep[1ES$~$1553+113,  $z \simeq 0.43$,][]{falomo90,sanchezhess2014,nicastronature18,2022MNRAS.509.4330D} but has been most systematically monitored in $\gamma$ rays by the Large Area Telescope (LAT) on board  the \fermi {\it Gamma-ray Space Telescope}, which was launched in 2008.

Early LAT observations raised suspicions of an almost sinusoidal modulation in the light curve of PG 1553+113, and a tentative identification of a 2.1 year periodic or possibly quasi-periodic oscillation (QPO) was described in \citet{A15}, A15 hereafter.
%
\begin{figure*}[hhhhttttthhh!!]
\hspace{-0.7cm}
\includegraphics[width=1.06\textwidth]{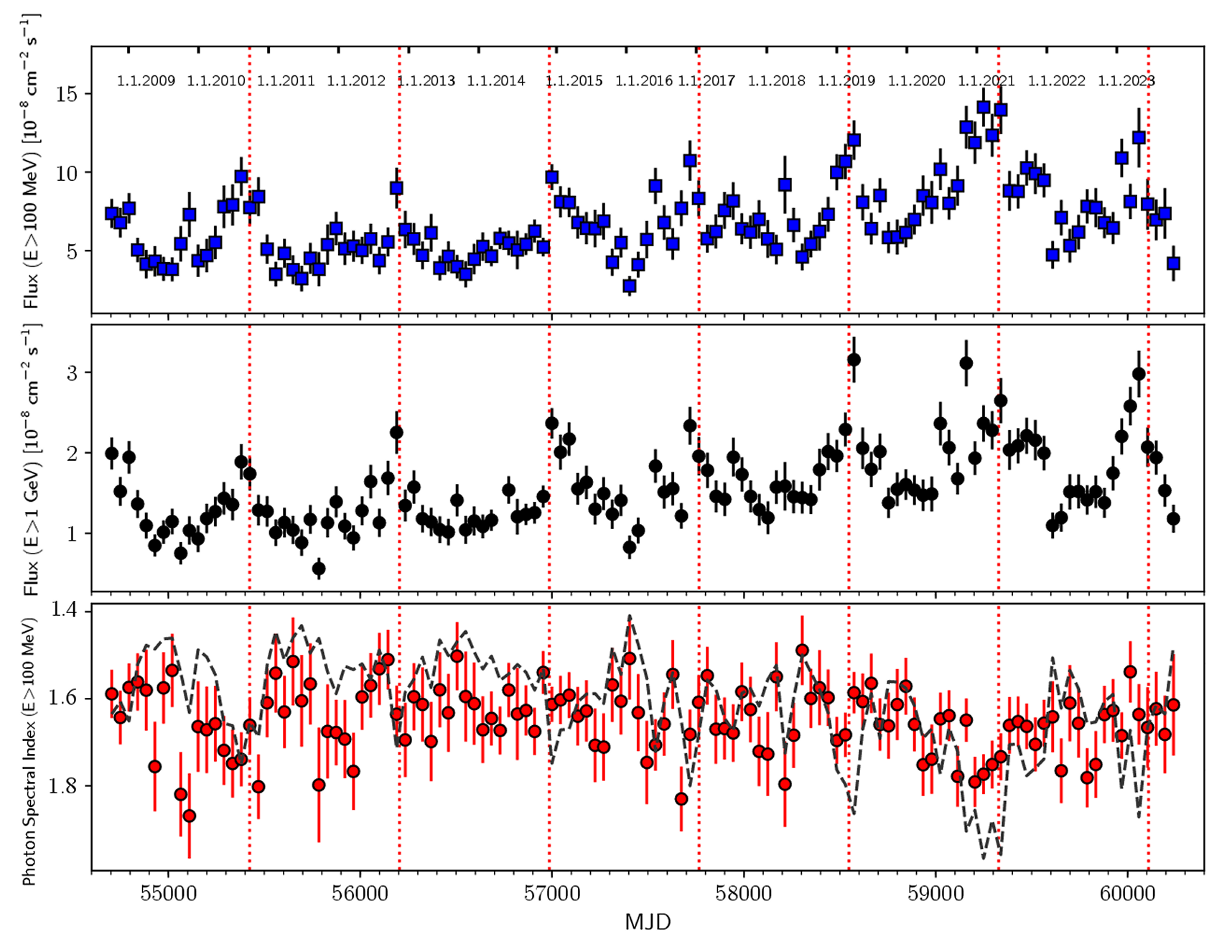}
\vskip -0.4cm
\caption{
\fermilat $\gamma$-ray integrated flux light curves and spectral index time profile of \pg\ over more than 15 years, from 2008 August to November 2023. The light curves above 100 MeV and above 1 GeV are shown with a constant 45-day time bin (first and middle panels); in the last panel we show the  \fermi\ LAT  spectral index ($E > 100$ MeV) as a function of time for the 45-day time bins, with a scaled version of the ($E > 100$ MeV) $\gamma$-ray  light curve superimposed as a dashed line.
}
\label{fig:latlc}
\vskip 0.3cm
\end{figure*}

A15 reported results based on LAT data from 2008 August 5 to 2015 July 19, MJD 54682.65 to MJD  57222.65 (a 6.9-year range), with radio and optical flux modulations in association with the $\gamma$-ray oscillation.

Several papers have followed up the results in A15, mainly corroborating the QPO discovery in $\gamma$-ray and optical bands with more data and more analyses \citep[for example ][]{cavaliere17,2019ApJ...875L..22C,sobacchi17,tavani2018,sandrinelli18,2020ApJ...895..122C,2020A&A...634A.120A,2020ApJ...896..134P,2024MNRAS.52710168P,2021A&A...645A.137A,2022JApA...43....9A,2023A&A...672A..86R,chen2024}.
Further blazars are also tentatively identified as producing correlated GeV $\gamma$-ray and optical QPOs behavior \citep[for example:][]{2020MNRAS.492.5524O,2021MNRAS.501...50S,2024MNRAS.52710168P,2024MNRAS.529.1365P}
.

The fact that \fermi\ LAT is producing long and essentially continuous light curves make these data ideal to use for searches for periodic or quasi-periodic variability. This, together with the indication of a periodic oscillation in \pg, has stimulated systematic studies of sources sampled from the \fermilat\  Third and Fourth Catalogs of Active Galactic Nuclei (3LAC, \citealt{3LAC} and 4LAC, \citealt{2020ApJ...892..105A})     \citep[for example, ][]{2016AJ....151...54S,itoh2016,2017MNRAS.471.3036P,sandrinelli18,2020A&A...634A.120A,2020ApJ...896..134P,2020ApJS..250....1T, 2024MNRAS.52710168P,2024MNRAS.529.1365P}.

The literature contains many studies, based on LAT and other data, with claims of QPO detections in variable blazar light curves. Many of these claims are controversial, primarily because blazars and other AGN exhibit erratic red-noise variability that can easily mimic periodicity, in short  time sequences containing few oscillation cycles \citep{corbet07,2009ApJ...691.1021D,2016MNRAS.461.3145V}. A common approach to avoid, or at least limit, this risk is to estimate period significance relative to simulated red noise. The present consensus is that the most sound simulation method available is the one described in \citet{2013MNRAS.433..907E}.
A claim of binary SMBH, with characteristics similar to \pg, was recently presented by \citet{oneill22}, for the BL Lac object PKS 2131-021, with a 4.8-year radio band flux periodicity (2.1 years in the rest frame).
Three brightest epochs in the LAT $\gamma$-ray light curve\footnote{\texttt{https://fermi.gsfc.nasa.gov/ssc/data/access\\/lat/LightCurveRepository/}} appear to precede the radio maxima by about a year.

In A15 we introduced a close ($0.005-0.01$ pc separation) gravitationally bound binary system of SMBHs (total mass $\sim 5\times 10^8\,\textrm{M}_{\odot}$) as a possible explanation for the observed periodicity. This would place the system in a state prior to the early inspiral, low-frequency, gravitational-wave (GW) driven regime, in agreement with \citet{2021MNRAS.506.1198D}. This state, with weak GW emission, would last  $t\sim10^{5}–10^{6}$ years, followed by a rapid inspiral and merger of the two black holes \citep{peters64,1983bhwd.book.....S}.

The scenario with a sub-parsec binary has been adopted or discussed in follow-up papers \citet{2017MNRAS.471.3036P,2020ApJ...895..122C,2018acps.confE..41C,2020ApJ...896..134P,2020A&A...634A.120A,2024MNRAS.52710168P}, generally with the $\gamma$-ray QPO confirmed at a $\sim 3-4\sigma$ level.

Evaluating the different proposed models for the QPO requires additional information, which might be provided by other types of observations. These include polarization and radio (parsec-scale) flux-structure long-term data as important probes of any regular wobbling \citep{caproni17,lico}, helical patterns and polarimetric rotations in the jet, and  potential identification of double-component spectroscopic line profiles \citep{graham15}.

In multi-epoch GHz milliarcsecond resolution maps by VLBA contained in the MOJAVE/2 cm Survey Data Archive \citep{lister09}, the radio core-dominated and magnetic-field-dominated BL Lac object \pg\ shows
a limb-brightened and diffuse-jet structure. These features might be the first indications of a jet with a precessing baseline, wobbling motion by torques of the inner disk, or alternatively an intrinsic pulsational accretion flow \citep{lister13,caproni17,lico}. Such geometric dynamics would cause periodic changes of the relativistic Doppler boosting, magnetic field stresses, magnetic reconnections and jet MHD instabilities \citep{cavaliere17,sobacchi17,tavani2018,2019ApJ...875L..22C,rieger19,lico}.

In this paper we describe in \S \ref{sect:data} the {\it Fermi} LAT $\gamma$-ray data, together with multi-wavelength X-ray ({\it Swift} XRT and {\it Rossi XTE} All Sky Monitor) data, optical (Tuorla, KAIT, CSS, {\it Swift} UVOT, ASAS-SN) data, and 15 GHz data from the OVRO radio-telescope.
In \S  \ref{sect:timing} we report multiple approaches in time-domain variability/periodicity analysis, cross-correlation and estimation of the $\gamma$-ray QPO coherence, with the findings suggesting a potential energy dependence. In \S  \ref{sect:discussion} we introduce some possible astrophysical interpretations, and in \S \ref{sect:conclusions} we summarize our findings. Appendix \ref{sect:appendixopticalhistorical} contains optical historical data from plates (years 1912 to 1988), and {\it Rossi XTE} ASM data (from year 1996 to 2011) for reference.

\section{Gamma-ray, X-ray, Optical and radio-band light curve data} \label{sect:data}

\subsection{{\it Fermi} LAT data}

The {\it Fermi} LAT
is a pair conversion  detector with a 2.4 sr field of view,  sensitive to $\gamma$-rays from $\sim 20$ MeV to $>300$ GeV \citep{atwood09}.
The present work uses the  Pass 8 LAT database \citep{atwood13}.
The {\it Fermi}-LAT operating mode allows it to cover the entire sky every two spacecraft orbits ($\sim$1.6-hour orbital period), providing a regular and uniform
view of  $\gamma$-ray sources, sampling timescales from hours to years.
This work uses observations of \pg~covering $\sim15.3$ years
(2008 August 4 to 2023 November 13, Modified Julian Day, MJD, 54682.65--60261.65).
The LAT data analysis employed the standard \texttt{ScienceTools} v11r01p01\footnote{\texttt{http://fermi.gsfc.nasa.gov/ssc/data/analysis\\/documentation/}} package, selecting events from 100 MeV$-$ 300 GeV 
with {\tt P8R3\_SOURCE$\_$V2} instrument response functions, in a circular Region of Interest (ROI) of 20$^\circ$ radius centered on the position of \pg\ (15h55m43.0440s +11d11m24.365s, J2000, \citealt{pos}).
We applied a zenith angle cut of $>90^\circ$ in order to reduce the contamination from the Earth limb and the standard data quality cuts (DATA\_QUAL $>$ 0 \&\& LAT\_CONFIG == 1) for the extraction of good time intervals. We selected only “SOURCE” event class (LAT EVENT\_CLASS=128) and event type “FRONT+BACK” (LAT EVENT\_TYPE = 3). We used files \texttt{gll\_iem\_v07} and \texttt{iso\_P8R3\_SOURCE\_V2\_v02} to model the Galactic and isotropic diffuse emission respectively.

A binned maximum likelihood model fit technique was applied to each time bin with a power-law spectral model and photon index  left free to vary  for \pg.
The background model included sources from the 4FGL catalog \citep{4FGL} within the ROI of 20$^{\circ}$; the fluxes and photon indices of background sources within a radius of 10$^{\circ}$ from the source of interest
were left free to vary.

The resulting light curves (LC), in different time and energy bins are shown in Figure \ref{fig:latlc}.

\begin{figure*}[htt!!]
\vskip -0.8cm
\hskip -1.1cm
\includegraphics[width=1.1\hsize]{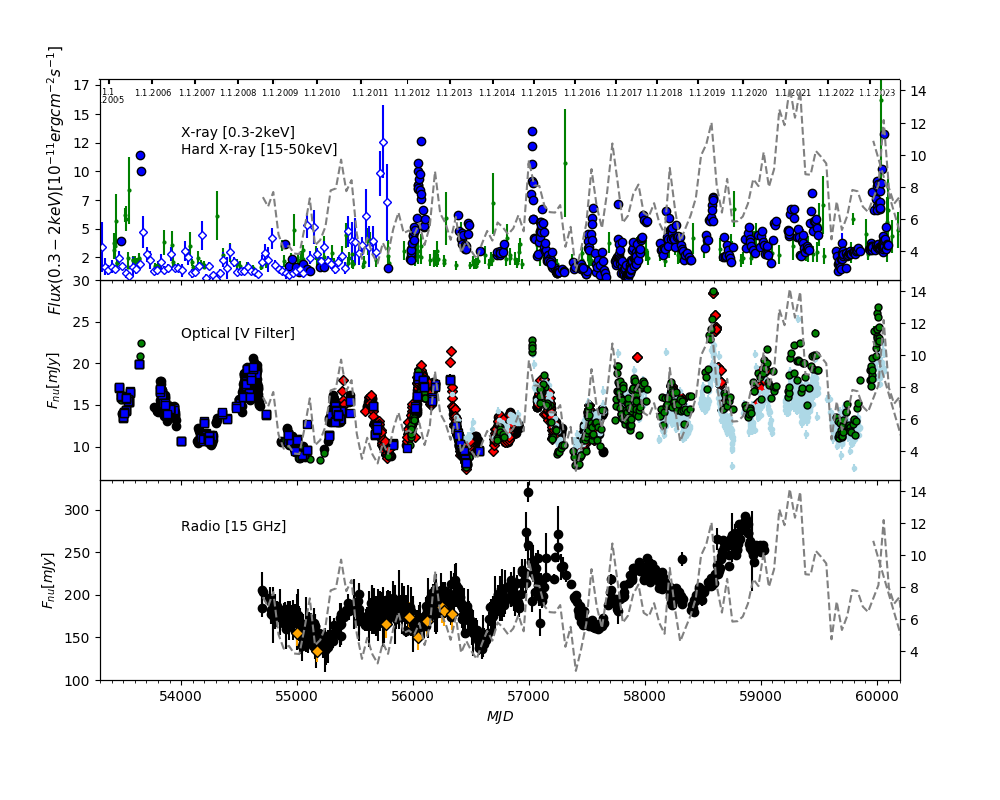}
\vskip -1.4cm
\caption{ %
Multi-wavelength light curves of \pg\ at X-ray and soft $\gamma$-ray, optical and radio bands with the LAT $\gamma$-ray light curve superimposed as a dashed gray line ($E > 100$ MeV). Top panel: \textit{Swift}-XRT integrated flux (0.3--2.0 keV) shown as blue filled circles,
and {\it RXTE} data points averaged into 30-day time bins  extrapolated to the 0.3--2.0 keV  energy flux (\textit{Swift}-XRT energy band) shown as  blue empty diamonds. \textit{Swift}-BAT count rate at 15-50 keV, multiplied by an arbitrary constant to scale it to the X-ray light curves, shown in green.
Central panel:  optical flux density from Tuorla (R filter, black filled circle points), Catalina CSS (V filter rescaled, blue filled squared points), KAIT (V filter rescaled, red filled diamond points), \textit{Swift}-UVOT (V filter rescaled, green filled circles) and
All-Sky Automated Survey for Supernovae (V filter rescaled, light blue filled  points).
Bottom panel: 15 GHz flux density from OVRO 40\,m (black filled circles) and parsec-scale 15 GHz flux density from VLBA (MOJAVE program, filled yellow diamonds). }
\label{fig:multifreqlc}
\end{figure*}

The $\gamma$-ray flux modulation oscillation is also found using different time bin sizes. In particular in the Appendix \ref{sect:3daybinlightcurveanalysis}  we report results from the the public data from the {\it Fermi}-LAT light curve repository at the FSSC \citep{kocevski21}.

\begin{figure*}[htt!!]
\hspace{-0.6cm}
\vspace{-0.5cm}
\resizebox{0.38\hsize}{!}{\rotatebox[]{0}{\includegraphics[align=c]{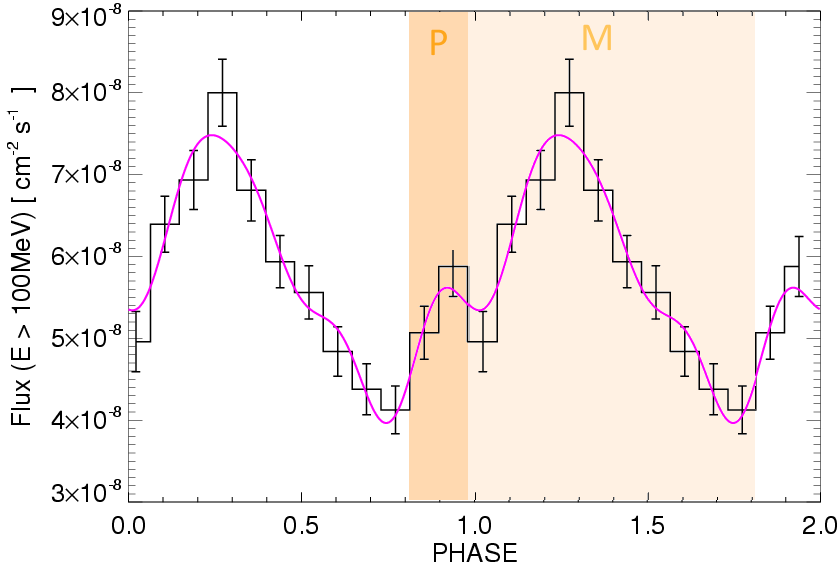}}}
\hspace{-0.2cm}
\resizebox{0.66\hsize}{!}{\rotatebox[]{0}{\includegraphics[align=c]{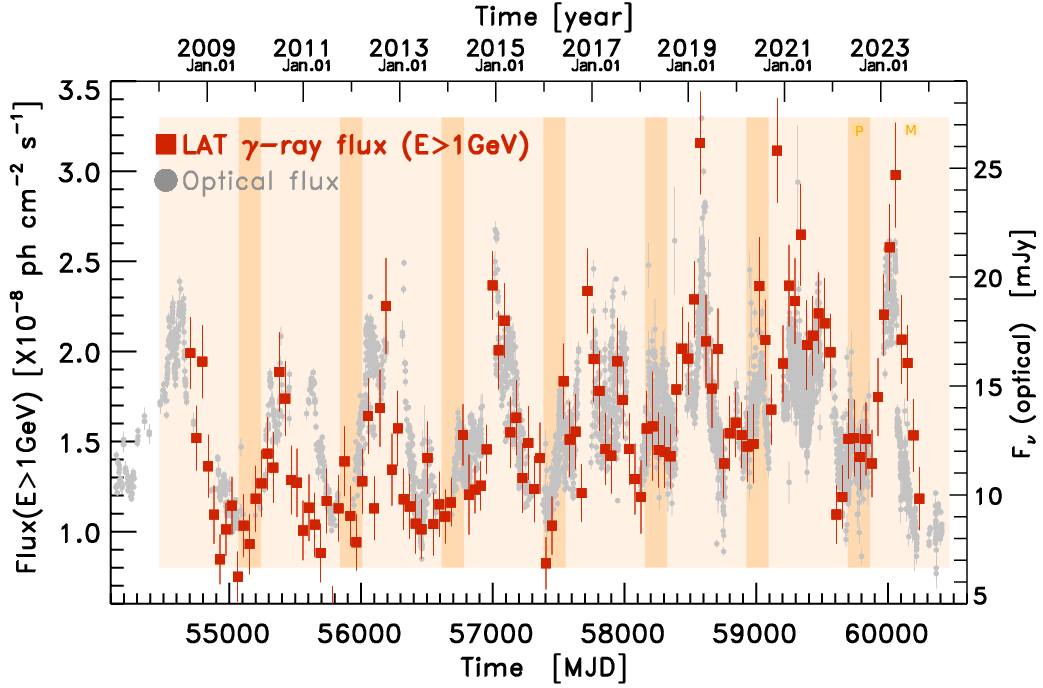}}}
\vspace{0.3cm}
\caption{
Left panel: \textit{Fermi}-LAT epoch-folded ($E > 100$ MeV) flux light curve with superimposed Fourier pulse fit, showing a slight bimodal peak. Right panel: superposed \textit{Fermi}-LAT ($E>1$GeV) and optical, rescaled, light curves for visual comparison (Tuorla, Catalina-CSS, KAIT,  Swift-UVOT). The periodic $\gamma$-ray Fourier oscillating pulse  precursor/maximum (P/M) epoch zones are also highlighted. These zones are taken to be 120 days long with start at MJD 57440 and period of 796.5 days}
\label{fig:precursor_phase}
\vskip 0.3cm
\end{figure*}
%

\subsection{{\it Neil Gehrels Swift} XRT and BAT data}
Four hundred thirty five, 435, {\it Swift} observations of \pg\ between 2005 April 20, and 2023 May 31, (MJD 53480.2--MJD 60095.8), are analyzed. The {\it Swift} X-Ray Telescope  (XRT) unabsorbed 0.3--2 keV flux light curve is presented in Figure \ref{fig:multifreqlc}. While all the PC-mode data were used, WT-mode data are omitted for the last 1.5 years of this data range.

XRT data were first calibrated and cleaned
({\tt xrtpipeline}, XRTDAS v.3.7.0), and energy spectra were extracted from a
region of 20 pixel ($\sim$47 $\arcsec$) radius, with a
nearby 20 pixel radius region for background.
Individual XRT spectra were well fitted with a log-parabolic model,
with neutral-hydrogen-equivalent column density fixed to the Galactic value of $3.6 \times 10^{20}$ cm$^{-2}$ \citep{kalberla05}.
For completeness we have also reported the historical public light curve
from {\it Swift} BAT. We depict only data points that have a relative uncertainty Flux/$\Delta $Flux $>$ 0.5. The
data are publicly provided by the {\it Swift} BAT team \citep{batbat} through
a dedicated webpage at at NASA Goddard Space Flight Center \footnote{https://swift.gsfc.nasa.gov/results/transients/weak/QSOB1553p113/}.

%


\subsection{{\it Rossi} XTE ASM data}
\label{xray-data}
X-ray flux densities in the 1.5--12 keV band were calculated from the public high-level archive \citep{1996ApJ...469L..33L} of the All-Sky Monitor (ASM) on board the {\it Rossi X-Ray Timing Explorer} satellite ({\it RXTE}) for \pg\ in the 15.7-year long dataset from 1996 Jan. 6, through 2011 Sep. 23, (MJD 50088--55827), using daily binning (see Appendix \ref{sect:3daybinlightcurveanalysis} for the complete 1-day bin count-rate time series). Data after MJD 55450 (last year of the light curve) are of poor quality because of the {\it RXTE} cathode loss.

Since ASM source X-ray counts were given as negative on many days (fluxes below the detection threshold), timing analysis via fast Fourier transform (FFT) and Lomb-Scargle periodogram (LSP) power spectra, was applied to both the original data and to data where we omitted negative counts and upper limits, using 1-day bins and other binnings as checks. The equivalence $1~\textrm{Crab~unit} = 75$ ASM count~s$^{-1}$ and the conversion $1\,
\textrm{Crab}~[2-10~\textrm{keV}] =
2.4\times10^{-8}~\textrm{erg}~\textrm{cm}^{-2}~\textrm{s}^{-1}$
are adopted (factor $2.79\times10^{-8}$ erg cm$^{-2}$ s$^{-1}$  to compare with the XRT $0.3-2$ keV fluxes in figure \ref{fig:multifreqlc}, evaluated from \citet{2005SPIE.5898...22K}).

\subsection{Optical, {\it Swift} UVOT, KAIT, Tuorla, Catalina CSS, ASAS-SN}
\label{optical-data}

Aperture photometry  for the UVOT V-band filter was performed.  The task is included in the HEASoft package
(v.6.23 by NASA HEASARC; 5\arcsec\, radius source aperture and two 18\arcsec\, apertures for the
background evaluation). Counts were converted
to fluxes using the standard zero points \citep{Breeveld}, and de-reddened using
the appropriate values of E(B--V) \citep{sf11} for the effective wavelengths of the UVOT filters. Tuorla blazar monitoring program \citep{takalo08} data are those published in A15, obtained with a 35-cm telescope attached to the Kungliga Vetenskapsakademi (KVA) telescope on La Palma and the Tuorla 1-m telescope in Finland
\citep[A15 and ][]{2018A&A...620A.185N}, with optical observations typically performed two to three times per week.

Public data from the Katzman Automatic Imaging Telescope (KAIT), the Catalina Sky Survey (CSS) and the All-Sky Automated Survey for Supernovae (ASAS) programs were also added. V-band magnitudes were scaled with an inter-calibrated fixed offset, (from an average of daily V--R)
to the R-band values.

Figure \ref{fig:multifreqlc} shows multi-frequency light curves. The light curves in  different wavelengths indicate a coherent periodic behavior from radio to $\gamma$ ray. This is true especially for the optical and $\gamma$-ray fluxes, also apparent in Figure \ref{fig:precursor_phase}.

\subsection{Radio OVRO data}

Figure \ref{fig:multifreqlc} shows flux density, long-term monitoring of \pg\ at 15 GHz obtained by the Owens Valley Radio Observatory (OVRO) 40-m radio telescope\footnote{\texttt{http://www.astro.caltech.edu/ovroblazars/}}, MOJAVE program,  with observation cadence of this blazar between 1 and 23 days since August 2008 \citep{2011ApJS..194...29R}.
A cooled receiver (3.0 GHz bandwidth, centered on 15.0 GHz and 2.5 GHz reception bandwidth) is placed at the prime focus of the 40-m radio telescope, with dual off-axis corrugated horn feed, projecting to Gaussian beams (157\arcsec\, FWHM) on the sky separated in azimuth by 12.95\arcmin, and with flux densities measured using azimuth double switching after peaking up on-source \citep{1989ApJ...346..566R}. The OVRO flux density data have a minimum uncertainty
of 4 mJy in 32 s of on-source integration, and a typical rms relative error of 3\%.
%

\section{Variability analysis}
\label{sect:timing}
%
The $\sim 2.1$ year period in \pg\ tentatively identified in A15 was based on data covering only $\sim$3 oscillation cycles. The light curves used in the present study are more than twice as long  which allows us to make several analysis improvements:
\begin{enumerate}
\item  Use a pulsation model built on the A15 data and compare it with the phase and frequency of a similar model built on the new data. Since the two data sets are independent we can estimate the probability that two oscillations by chance will have a particular phase and frequency.
\item  Make a more strict significance estimate for the periodic oscillation in comparison with simulated stochastic red noise.
\item Perform a wavelet time-frequency study of pulsation evolution along time, and significance in relation to a theoretical stochastic (random-walk) first-order autoregressive, AR(1), model.
\item  Improve on earlier estimates of the correlation between the optical, radio and $\gamma$-ray variations in \pg.
\item  Set limits on the coherence of the oscillation.
\item  Study the energy dependence of the $\gamma$-ray quasi periodic oscillation (QPO).
\end{enumerate}
Below we describe each of these topics and present the associated results.

%
\subsection{Testing the periodicity hypothesis with new data and PDS pulse analysis}\label{sect:testing}
\begin{figure}[hhhhhhtt!!]
\vskip -0.6cm
\hskip -0.8cm
\resizebox{1.25\hsize}{!}{\rotatebox[]{0}{\includegraphics{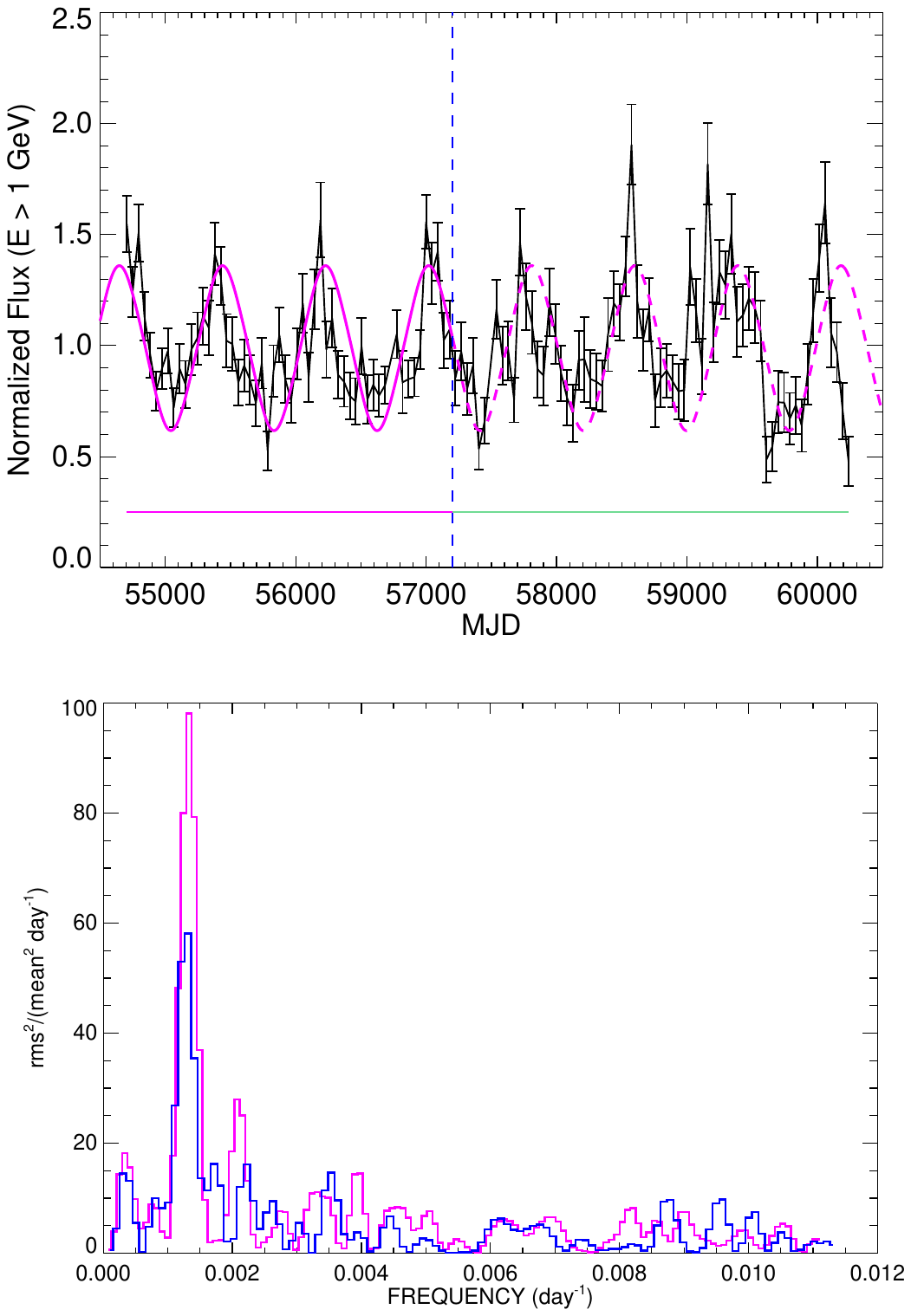}}}
\vskip -1.1cm
\caption{Top panel: A Fourier fit to the linearly detrended 1 GeV light curve up to MJD 57203, the data used in A15. The fitted oscillation is extrapolated and shown as a dashed line after MJD 57203 for comparison with the new data. Lower panel: Comparison of Power Density Spectra, PDSs, for the two independent data segments (blue: t$<$MJD 57203, purple: t$>$MJD 57203).}
\label{fig:sl1lc}
\end{figure}

The analysis in A15 showed that the Fermi-LAT light curve for the first 6.9 years could be described as a periodic modulation. By assuming a strictly periodic modulation and estimating its parameters from the data in A15 it is now possible to test that hypothesis by comparison with an independent fit of period and phase to the new data. In the present analysis we have done so by fitting a Fourier pulse (fundamental period plus 3 overtones) to the light curve ($E > 1$ GeV) of the first $\sim 7$ years and then extrapolating that modulation to later times. The energy selection and fitting procedure are identical to the corresponding analysis in A15. To avoid biasing the hypothesis test we tried no other procedures. Before the analysis, the full light curve was detrended by subtracting a least square fitted line. The light curve was then normalized to mean 1 and the same fractional variability (rms/flux) as the original light curve. The normalized light curve is shown in the upper part of Figure \ref{fig:sl1lc}. The resulting Fourier fit is also shown in the figure. The dashed vertical line at MJD 57203 indicates the separation between the data that were used for the Fourier fit (corresponding to A15) and the new data. The phase difference between the two oscillations is estimated by a cross correlation between the extrapolated Fourier template and the new data after MJD 57203. The estimated  time lag in phase units of the template with respect to the new data is $ 0.117 \pm 0.033 $, where the error value only takes the statistical uncertainties of the new data into account. The total uncertainty of the phase difference is dominated by the model extrapolation from the first observing epoch. The uncertainty range for the extrapolated phase at the midpoint of the new data is between $+0.094$ and $-0.062$. We conclude that the observed phase difference between the predicted and observed phase in the new data is marginally consistent when phase uncertainties are taken into account.

The oscillation periods for the two data segments, before and after MJD 57203, can be obtained from the Power Density Spectra (PDSs). These are shown in the lower part of Figure \ref{fig:sl1lc}. Each PDS is oversampled by a factor of 4, which matches the frequency uncertainty in our Fourier fit analysis. The uncertainty range is approximately two thirds of the width of the frequency bins. The difference in peak frequency for the two segments is approximately $5 \times  10^{-5}$ day$^{-1}$ which is less than the widths of the frequency bin ($1.01 \times 10^{-4}$ and $0.83 \times  10^{-4}$ day$^{-1}$ respectively). Assuming that the period in the new data could fall in any of the 118 frequency bins (neglecting the first 4 due to the linear detrending of the light curve), the chance probability for the two oscillations to be so close in frequency is 0.0085. The similarity in pulsation phase described above gives additional weight to the confirmation of the pulsation model for the previous data. A conservative conclusion is that the chance probability for the pulsation period and phase of the new independent data to be  close to those of the earlier data is less than 0.01.

\subsection{Estimating QPO significance with pure-noise light curve
simulations}\label{sect:emmanlcsimulations}

Periodicity studies of astronomical time series deal not only with the estimate of the period and its uncertainty, but with whether there is a (quasi)-regular periodicity or not. The statistical significance for such periodicity must then be defined.
This calculation is not trivial: significance can be estimated using the False Alarm Probability (FAP) \citep{2018ApJS..236...16V}, which measures the probability that the data are dominated by white/colored noise would lead to a signal peak in the periodogram of higher power than the   peak due to a truly periodic signal. The FAP can be estimated through light curve simulations and
the LSP \citep{1976Ap&SS..39..447L,1982ApJ...263..835S}, as the ratio $n/N$: the number $n$ of simulated time series with power peaks higher than the original light curve peak found, over the total number  of simulations ($N=10^6$ in our case).
The FAP is not a direct estimate of the periodicity significance, but rather the probability of a false alarm, which is substantially different. Simulations are randomizations of the original, true, light curve data. However the brown/red/pink, colored, noise power spectra of true light curves of highly variable blazars (and many other astronomical source types) tend to result in overestimated, not reliable, periodicity significances \citep{2013MNRAS.433..907E,2016MNRAS.461.3145V}.

In order to simulate pure-noise data with the same characteristics as the true $\gamma$-ray light curves of \pg, we used the Emmanoulopoulos algorithm \citep{2013MNRAS.433..907E}, to generate surrogate simulated light curves with  PDF and Power Density Spectra (PDS) similar to the true data light curve.
The resulting $n/N$ FAP value, evaluated as the greatest LSP power peak of each simulated light curve, is difficult to interpret;  therefore a corresponding significance, equivalent to the number of standard deviations $\sigma$ calculated from the cumulative of a normal distribution up to $1-n/N$, is introduced. If none of the simulations has a greater peak than the original peak ($n = 0$), a lower limit on the FAP must be set. For \pg\ we choose a lower limit of $1/N$, and a corresponding upper limit on the value of $\sigma$ is set too.

\begin{figure}[ttthhh!!]
\centering 
\hskip -0.9cm
\resizebox{1.1\hsize}{!}{\rotatebox[]{0}{\includegraphics{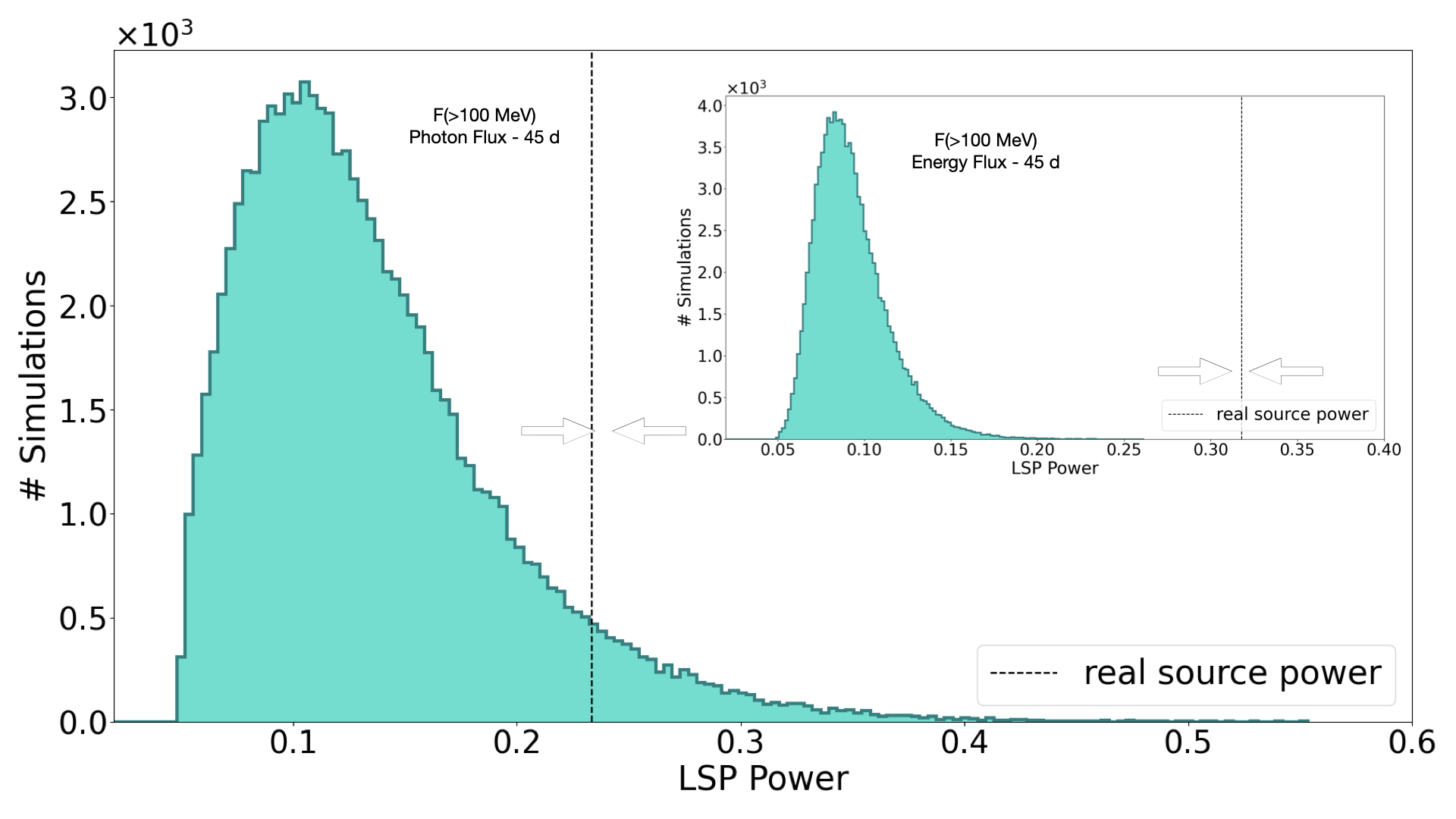}}}
\vskip -0.2cm
\caption{
Signal power with the Lomb Scargle periodogram (LSP) peak distributions of $10^6$ simulated light curves,  for the  45-day bin ($E>100$ MeV) energy flux light curve in the inset and for the  45-day bin ($E>100$ MeV) photon flux light curve in the main panel.
Dashed lines are the LSP power for the two, true data, light curves. The significance of the period here is $>4\sigma$ and $1.7\sigma$ for the energy flux  and photon flux light curves respectively.
}
\label{fig:45day20daybinLC_emmansimulDistrib}
\vskip 0.3cm
\end{figure}

\begin{table}[t!!!!]
\begin{center}
\footnotesize
\vskip -0.2cm
\caption{Summary of the $\gamma$-ray QPO significance in $\sigma$ for the various LAT light curves of PG 1553+113 presented or discussed in this work. \label{tab:sigmas}}\vskip -0.3cm
\begin{tabular}{l|c|c|c|c|c}
\tableline\tableline
$\gamma$-ray band  & \multicolumn{4}{c|}{$E>100$ MeV}  & $E>1$ GeV \\
\tableline
LC type & \multicolumn{2}{c|}{Photon flux} & \multicolumn{2}{c|}{Energy flux} & Photon flux   \\
\tableline
Bin size & 20 days    & 45 days    & 20 days    & 45 days    & 45 days   \\
LC original & $1.5\sigma$  & $1.7\sigma$  & $3\sigma$  & $>4\sigma$   & $3.5\sigma$   \\
LC detrended & $2\sigma$  & $2.3\sigma$  & $3\sigma$  & $4\sigma$  & $4\sigma$ \\
\tableline\tableline
\end{tabular}
%
%
%
%
\normalsize
\end{center}
\vskip -0.6cm
\end{table}
%
We apply this approach for the different $\gamma$-ray light curves for \pg\ presented in this work (Table \ref{tab:sigmas}): photon/energy flux, fluxes for ($E > 1 $ GeV) and ($E > 100 $ MeV), 45 and 20 day time bin light curves (also 3-day bin data reported in Appendix \ref{sect:3daybinlightcurveanalysis}). For each we calculate $10^6$ simulations.
The $\gamma$-ray light curves have a hint of a secular increase in flux, and de-trended light curves (and their proper simulations) are also evaluated for comparison.
The results are reported in Table \ref{tab:sigmas}. The
distributions for both of the simulated data for the true 45-day  photon and energy flux light curves, are shown as example in Figure \ref{fig:45day20daybinLC_emmansimulDistrib}.

\subsection{Wavelet time-frequency study of the QPO signal's evolution}\label{sect:wavelets}

\begin{figure*}[htt!!]
\hskip -0.5cm
\includegraphics[width=1.05\hsize, angle=0]{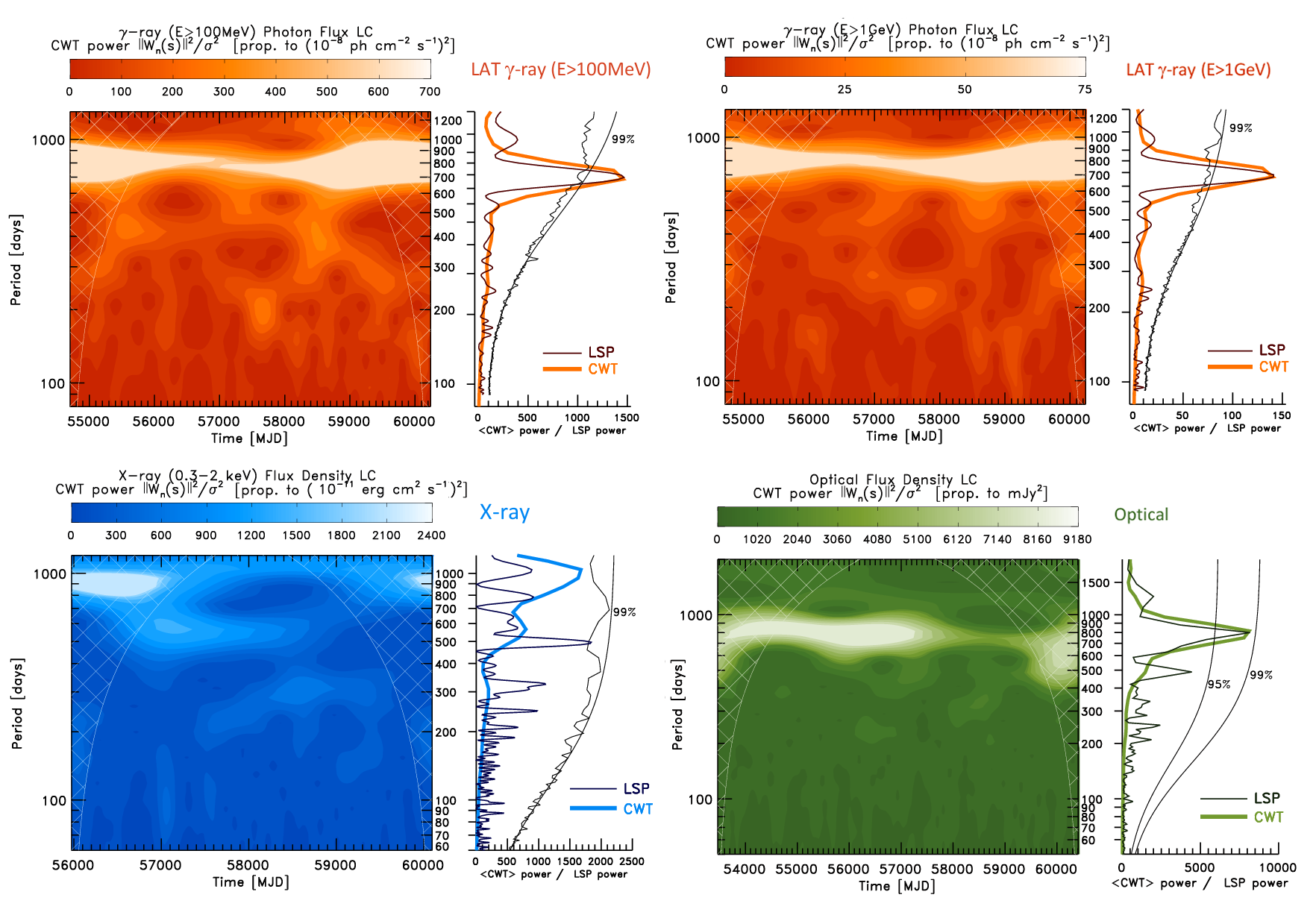}%
\vspace{-0.4cm}
%
\caption{Top panels: 2D image filled (orange color scale) contour plots of the Morlet CWT $\gamma$-ray power spectrum, the ``scalogram'', respectively for the \fermilat\ $E>100$ MeV and $E>1$ GeV, 45-day bin, photon flux light curves of Fig. \ref{fig:latlc}. Side plots are the global CWT power spectrum averaged along the MJD epochs, and the one-dimension LSP power spectrum in the bias-corrected REDFIT implementation. The LSP power axis is rescaled to the scale of the global-CWT comparing the relative main bumps. The 99\% false alarm confidence level lines (parametric $\chi^2$ and Monte Carlo) against the null hypothesis of a origin by pure  stochastic AR(1)-noise process realization, are shown.
Lower panels: 2D scalogram filled color scale contour plot of the Morlet CWT  (blue color scale) of the \textit{Swift}-XRT integrated (0.3-2.0 keV) flux density (omitting the few, too sparse, XRT data prior to the year 2012), and 2D scalogram filled color scale contour plot of the Morlet CWT (green color scale) of the optical flux density, corresponding to light curves presented in Fig. \ref{fig:multifreqlc}.
Cross-hatched regions in all the panels are the ``cone of influence'', where spurious edge effects caused by finite time-series boundaries become important. Side plots are the corresponding global CWT and the LSP power with their false alarm confidence levels lines, scaled to the global CWT peak.
}
\label{fig:waveletandperiodgram}
\end{figure*}

In a generalized Fourier space, the continuous wavelet transform (CWT) is  able to both identify the dominant modes of variability and how those modes vary in time. CWT can also be quite sensitive to signals that would  go undetected by other methodologies \citep[e.g. ][]{mallatbook}. In Figure \ref{fig:waveletandperiodgram} we report the diffuse two-dimensional, time-frequency $t$ versus $f=1/T$, ``scalogram'' plots, with the normalized 2D modulus of the CWT energy density function $\left\| W_{n}(s) \right\|^{2}/\sigma^{2}$, where $1/\sigma^{2}$ measures the power relative
to white noise, corresponding to the LAT $E>1$ GeV and $E>100$ MeV $\gamma$-ray, XRT X-ray, and optical light curves of Figures \ref{fig:latlc} and \ref{fig:multifreqlc}. The Morlet mother waveform, a wave packet of a plane wave modulated by a Gaussian, is used here as the best tradeoff between epoch localization in $t$ and resolution in frequency $f$ (or equally period/periodicity timescale $T$). Lighter colors describe higher peaks, bumps and regions, and hence  stronger values of the 2D CWT power spectrum.
%
%

%
The orange filled-color scale 2D image contour plots in the top panel of Figure \ref{fig:waveletandperiodgram}, result from the Morlet CWT $\gamma$-ray power spectrum for the $E>100$ MeV and $E>1$ GeV, 45-day bin, light curves reported in Fig. \ref{fig:latlc}. Side plots show the global CWT power spectrum (i.e., averaged along the light curve MJD epochs on the x-axis) and the Lomb-Scargle periodogram (LSP) power spectrum, in the bias-corrected implementation given by the REDFIT method and software \citep{2002CG.....28..421S}, which also provides a mathematical first-order autoregressive AR(1)-model fitting, adequate for stochastic natural and astronomical processes. The method works directly on unevenly binned/gapped data, avoiding the introduction of bias during interpolation. Also we did not manipulate the light curve data, avoiding any linear detrending prior to analysis, even if this could have given us slightly higher significance values. With REDFIT we applied a Welch-window overlapped segment averaging, choosing to split the timeseries into to two segments that overlap by 50\% to reduce noise,  while avoiding too great a reduction of the spectral resolution, and then averaging their power spectra. We obtained the same results with rectangular and Welch window functions.
The REDFIT software also
calculates the significance levels on the LSP spectrogram based on parametric approximation with the $\chi^2$ calculated with respect to the computed AR(1)-noise spectrum (the null hypothesis), where the degrees of freedom depend on the number of data points. It also calculates the significance level with one thousand Monte Carlo random realizations of the AR(1) random-walk, a.k.a. red/brown/Brownian, noise model.

In general, any peak of the one-dimensional LSP spectrogram power above the 99\% confidence curve level, has less than a 1\% chance of being a false alarm product of stochastic red/Brownian AR(1) process realizations. In this case the parametric $\chi^2$ smooth and MonteCarlo AR(1) irregular confidence curve levels reported in the four side-panels in Figure \ref{fig:waveletandperiodgram} have similar trends and intensity.

Both the $>100$ MeV and $>1$ GeV panels in Fig. \ref{fig:waveletandperiodgram} show a coherent $\gamma$-ray global-CWT peak centered  at $\simeq 780$ days (2.1 years) with a significance of $>3\sigma$, which approaches $4\sigma$ for the 1 GeV flux light curve, as shown by the prominence of both of the bumps with respect to the  99\% AR(1)-noise false alarm confidence levels. Most of the signal power during this 15.3-year $\gamma$-ray light curve epochs is manifestly placed around this timescale, well represented by the clear/white horizontal strip in both of the CWT 2D orange-scale image plots at the top, with such strips sufficiently outside the, spurious, ``cone of influence'' region.  The CWT indicates therefore a dominant 2.1-year QPO $\gamma$-ray flux modulation, along the light curve time range.

In Figure \ref{fig:waveletandperiodgram}, the 2D Morlet CWT (blue color scale) plot of the \textit{Swift}-XRT integrated (0.3-2.0 keV) flux from MJD 55973, shows erratic variability and no significant QPO, i.e. a behavior substantially different from the GeV gamma-ray, optical and radio periodic modulation.

The LSP of X-ray data indicates a $<2\sigma$ potential peak around $\sim 498$-day timescale ($\sim 1.4$ years), not corroborated by the wavelet. The CWT has some X-ray signal in a region between $\sim 1.4$ and $\sim 1.8$ years, extended at least from June 2013 to Oct. 2020. Two high-power white ovals ($\sim 900-1000$-day timescale) are both placed at the edges of the X-ray light curve, when relevant X-ray outbursts occurred. In a such cross-hatched region, spurious power by edge effects to the CWT scalogram, is important. A stronger hint is recently reported in \citet{2024A&A...686A.300A}.

The 2D CWT (green color scale) plot of the 19.7-year optical light curve, has well-defined global-CWT and LSP peaks at $\simeq 790$ days ($\simeq 2.16$ years),  quite similar to the $\gamma$-ray QPO, and a rather stable evolution in time along the x-axis of the 2D scalogram image . The significance is lower ($\sim2\sigma$) as shown by the 99\% and 95\% false alarm confidence levels. This supports a multi-band QPO, also because the optical-$\gamma$-ray cross correlation, even if such wavelet and LSP of the optical data, when taken standalone, can still be  consistent with erratic red/brown-noise variability. We remember that, already in A15, we reported an apparent $754\pm20$ day periodicity in the 9.9-year range (MJD 53479-57110) for the optical data.

In this work the wavelet analysis for the 15 GHz radio light curve is not presented, since the date range is similar to that considered in A15, and the results are similar.

The results of the wavelet analysis and the REDFIT spectrogram ($>3\sigma$-significance $\gamma$-ray QPO, $2\sigma$-significance optical QPO), although insufficient to claim a deterministic periodicity, raise a more solid possibility of there being a true QPO, when supported, as in this work, by the other methods which we illustrated in the previous sections.

\begin{figure}[hhhhhttttt!!]
\centering 
\hspace{-1.0cm}
\resizebox{\hsize}{!}{\rotatebox[]{0}{\includegraphics{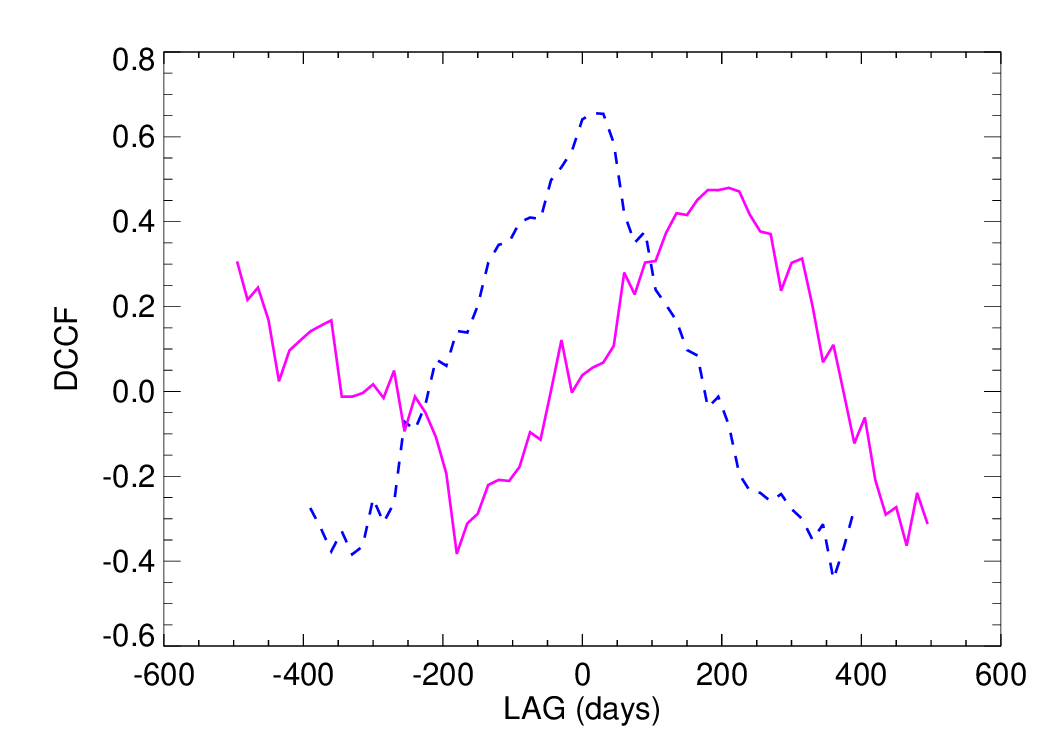}}}
\vskip -0.4cm
\caption{
Discrete Cross Correlation Functions (DCCFs) between \textit{Fermi}-LAT ($E > 1$ GeV) and the rebinned optical and radio (15 GHz) light curves. Blue (dashed) curve: Gamma versus optical. Purple (solid) curve: Gamma versus radio. Estimated time lags relative to $\gamma$ rays are $6 \pm 18$ days for the optical and $188 \pm 28$ days for the radio.
}
\vskip -0.4cm
\label{fig:DCCF_gamma_vs_optical_and_radio}
\end{figure}

\begin{figure}[hhhhhhtt!!]
\hspace{-1.0cm}
\resizebox{1.16\hsize}{!}{\rotatebox[]{0}{\includegraphics{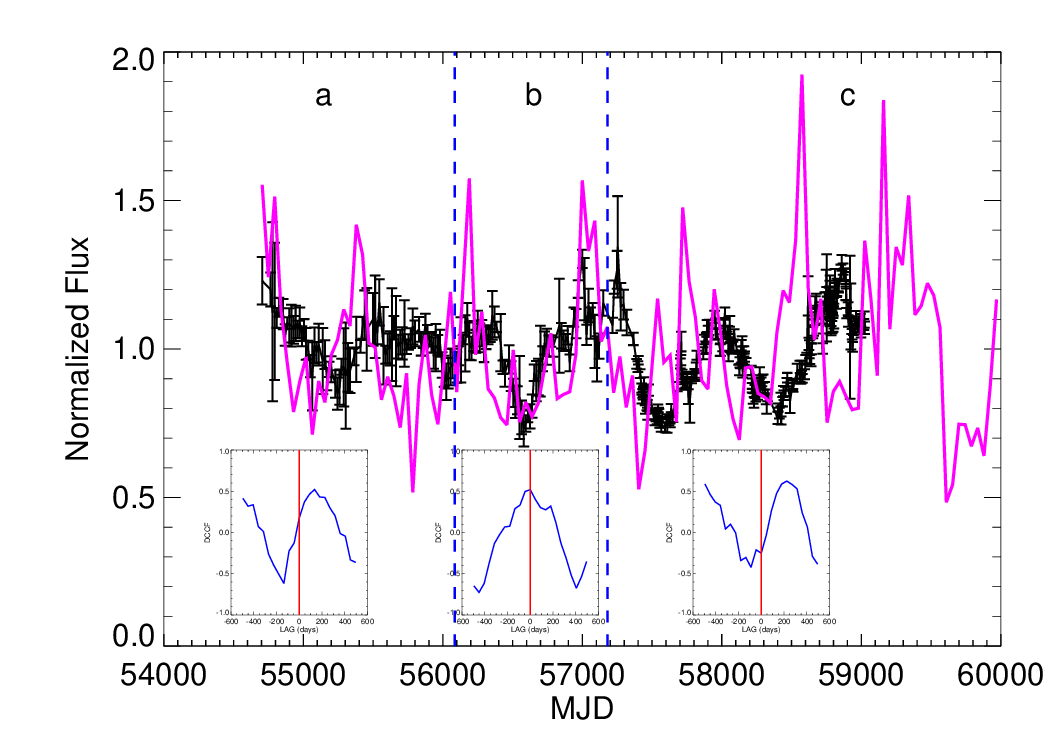}}}
\vskip -0.5cm
\caption{OVRO 15 GHz light curve (in black with error bars) together with LAT ($E > 1$ GeV) flux (purple). Inset panels show cross correlations (DCCF) between the light curves of segments a--c. While the overall time lag of the 15 GHz flux relative to $\gamma$-ray is on the order of 200 days there are substantial cycle-to-cycle variations. Lag 0 in the DCCF plots is indicated by vertical lines.}
\label{fig:LATOVROdccfandlcs}
\vskip -0.3cm
\end{figure}

\subsection{Gamma-ray - optical and radio correlation}
\label{sect:correlations}

The relationship between the $\gamma$-ray, optical and radio variability in \pg\ was investigated in A15, using the first 6.9-years of \textit{Fermi}-LAT observations. The time lag between the optical and the \textit{Fermi}-LAT ($E > 1$ GeV) light curves, both 20-day binned, was found to be $10 \pm 51$ days, with the optical leading but still consistent with zero lag. The relation between the $\gamma$-ray and  radio was found to be more complex with the 15 GHz flux lagging the $\gamma$-ray flux by some 2-3 months.

Using the updated light curves in Figure \ref{fig:multifreqlc}, we have calculated the discrete cross correlation function (DCCF) for \textit{Fermi}-LAT ($E > 1$ GeV) with respect to optical and OVRO (15 GHz) flux. The optical R and V-band observations shown in Figure \ref{fig:multifreqlc} were normalized and merged to a single optical light curve. The optical measurements are very irregularly distributed in time and were therefore averaged into 20-day bins before the analysis. Since early OVRO measurements had lower signal-to-noise than later observations, the data up to MJD 57200 were sampled in 15-day bins (with typically 3 original points per bin). The cross correlations are shown in Figure \ref{fig:DCCF_gamma_vs_optical_and_radio}. Time lags were estimated by fitting a Gaussian function to the DCCF peak. Lag uncertainties were estimated by a Monte Carlo method (bootstrap resampling and adding white noise equal to total standard deviation; \citep[see ][for details]{larsson12}. The estimated time lags relative to the $\gamma$-ray are $6 \pm 18$ days for the optical and $188 \pm 28$ days for radio.
While the $\gamma$-ray-optical correlation is consistent with zero lag, the large 15 GHz time lag is not unusual for blazars  \citep{fuhrmann14,maxmoerbeck14}. It is also clear that the radio light curve is less regular than the one in $\gamma$ rays. This is true for the first part of the OVRO light curve, segments a and b in Figure \ref{fig:LATOVROdccfandlcs}. By contrast, the light curve after approximately MJD 57200 (segment c) shows a much more clean, quasi-sinusoidal modulation with a visually apparent time lag relative to the $\gamma$-ray flux. This is corroborated by the cross correlation for that part of the data. A fit to the DCCF in the lower right of Figure \ref{fig:LATOVROdccfandlcs} gives an estimated lag $= 226 \pm 25$ days. The division into time intervals a-c here is only meant to illustrate what seems like a variation in correlation behavior between the radio and $\gamma$-ray bands.

\begin{figure}[bbbbbb!!]
\centering 
\resizebox{\hsize}{!}{\rotatebox[]{0}{\includegraphics{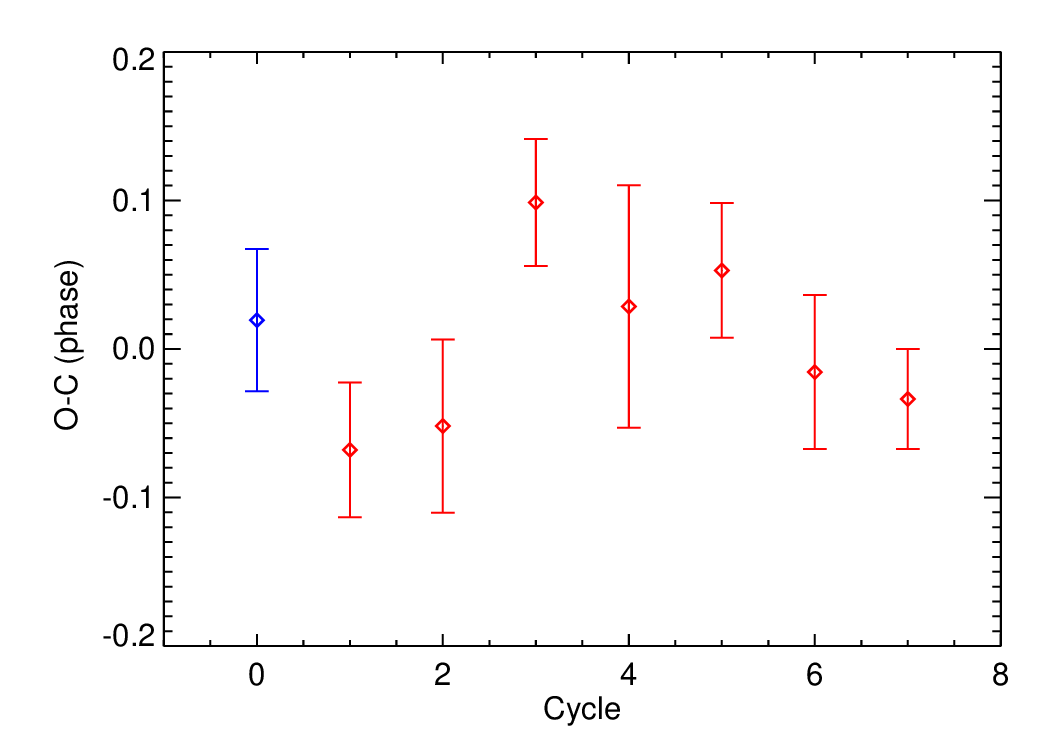}}}
\vskip -0.5cm
\caption{Pulse phase variation relative to a strictly coherent period. The pulse phase for each cycle is estimated by cross-correlation with a Fourier model fitted to the full light curve. The first point is from the optical data in figure \ref{fig:multifreqlc}. All other points are from the $E > 1$ GeV light curve. Cycle number correspond to N in Equation \ref{eq:ephem}. A observed-computed (O-C) diagram compares the observed time of maximum flux O with the calculated time C, assuming a known constant period $T$, and is a classical method for studies of period changes in variable stars.}
\label{fig:sl1oc}
\vskip -0.3cm
\end{figure}

\subsection{Coherency} \label{sect:coherency}

The question of whether the periodicity in \pg\ is quasi-periodic or strictly coherent has important implications for the interpretation and theoretical modeling of this periodicity.
The comparison of pulse phases for the two segments made in Section \ref{sect:testing} implies an absolute phase drift of $ \lesssim 0.12$ over a few cycles. It is also consistent with a strictly coherent periodicity. Since the presently available $\gamma$-ray data cover almost seven pulse cycles it is also possible to look for any significant phase jitters over this epoch.
The times of individual pulse peaks were estimated by cross correlating one cycle of \textit{Fermi}-LAT data centered on the peak with a Fourier pulse. Our timings of the seven peaks were based on the \textit{Fermi}-LAT ($E > 1$ GeV) light curve and one peak timing from the optical observations preceding the \textit{Fermi}-LAT data.
Combining the optical and $\gamma$-ray timings is motivated by the strong correlation with small or negligible time lag between the two bands as shown in Section \ref{sect:correlations}.  A linear fit to the  timing of the $\gamma$-ray peaks results in the following ephemeris, where $N$ is the cycle number,

\begin{equation}
\label{eq:ephem}
T_{peak}= {\rm MJD}\,54659 (\pm 31) + 771.5 (\pm 7) \times N\,{\rm days}.
\end{equation}

The value $771.5\pm7$ (i.e. 2.11 years) is the modulation period.
Deviations relative to this ephemeris are shown in Figure \ref{fig:sl1oc}. The observed phase jitter is slightly larger than expected for a strictly coherent periodicity ($\chi^2 = 11 $ for 6 d.o.f., probability = 0.09). We can not exclude that this is an effect of a slight underestimation of the data errors.

\subsection{Energy dependence of the $\gamma$-ray QPO}
\begin{figure}[hhhtt!!]
\hskip -1.2cm
\centering 
\vskip -0.2cm
\hskip -0.5cm
\resizebox{\hsize}{!}{\rotatebox[]{0}{\includegraphics[width=1.2\hsize]{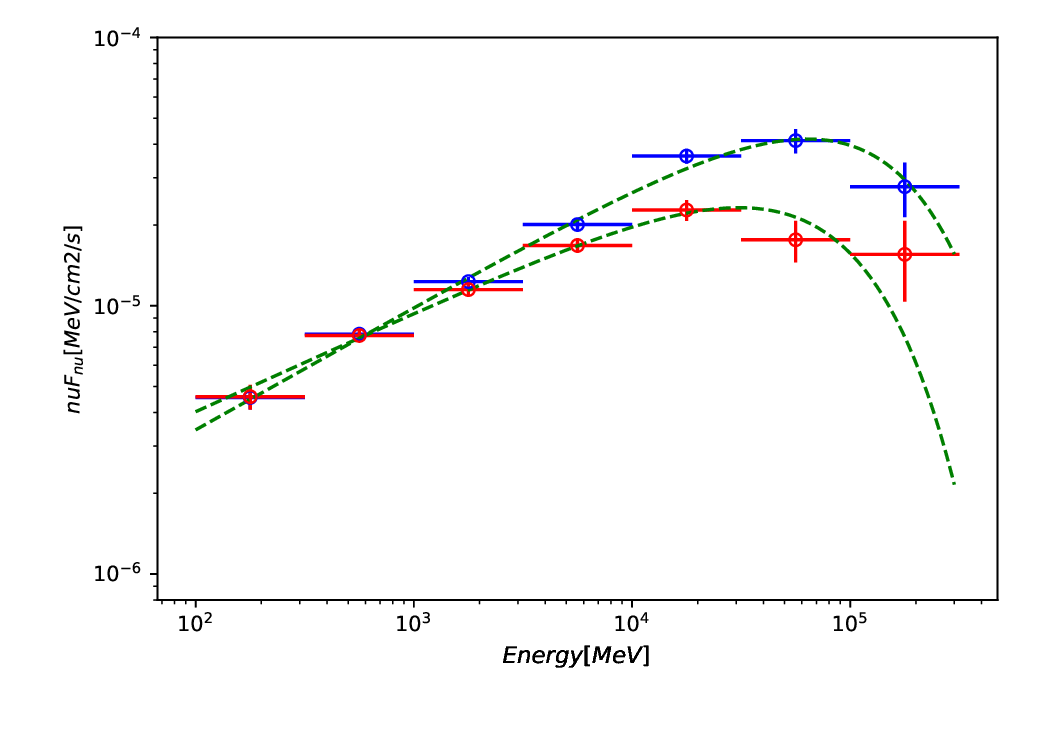}}}\\[-0.2cm]
\caption{
Spectral energy distribution of  \pg ~in the precursor phase shown in red (grey) circles as highlighted in the right panel of Fig. \ref{fig:precursor_phase} and the main phase shown in blue (dark) circles. A cutoff power-law function was used, shown by the green dashed lines.
}
\label{fig:sed}
\end{figure}

The spectral energy distributions (SEDs) in different phase ranges of the periodic variations, were extracted with the binned maximum likelihood method with the same analysis parameters as were adopted for LC extraction. Each of the spectral bins were obtained using
a power-law model for the source of interest with a spectral index fixed to 2.
We selected data in two different phase ranges, selecting the substructure in the pulse shape with four Fourier components as shown in Figure \ref{fig:precursor_phase} (first panel).
We estimated the SED for the precursor phase interval of about 0.5--0.7 in Figure \ref{fig:precursor_phase}. Note that these phases are based on a slightly different period than that given in Equation \ref{eq:ephem}. The precursor time intervals are highlighted in Figure \ref{fig:precursor_phase} (second panel). For comparison we also extracted  the SED in the main part of the peak with a complementary time selection (phase interval of about 0.7--1.5). The two SEDs are shown in Figure \ref{fig:sed}.
We fitted the SEDs with a cutoff power-law with best fit spectral index values of 1.62 $\pm$0.03 (${\chi}^2_{red}$ of 1.2 ) and 1.51$\pm$0.02 (${\chi}^2_{red}$ of 1.2 ) and a energy cutoff equal to 84$\pm$23 GeV and 140$\pm$30 GeV for the precursor and main peak phases, respectively.
The precursor phase shows a softening  in the spectral index with respect to the main part of the periodic modulation with a significance of about 3 $\sigma$.
This behavior is corroborated by  the spectral index  evolution  as a function of time using the  45-day  time  bins of $\gamma$-ray light curves as reported in Figure \ref{fig:latlc}. A few stochastic  flaring activities show a softening of the spectrum during the precursor time frames.
Moreover we also performed periodicity searches   on the energy flux light curve, and using this different set of data we found  a higher significance as reported in Section  \ref{sect:emmanlcsimulations},  confirming the hypothesis that the periodic modulation is more prominent at higher energies.

\section{Discussion} \label{sect:discussion}

\subsection{Candidate astrophysical scenarios for periodicity} \label{sect:introduscuission}
The fairly long-lived $\sim 2.1$-year period and correlated $\gamma$-ray and optical cyclic modulation in \pg, has been discovered thanks to the $\sim15$-year continuous all-sky survey performed by \textit{Fermi}-LAT, with no degradation of its performance.

Advection-dominated accretion regimes (magnetically-dominated/-arrested accretion flows: MDAF and MAAF) related to QPOs, could explain turbulent, peculiar radio kinematics related to radiatively inefficient, high-energy, TeV, BL Lac objects like \pg\  \citep{fragile09,karouzos12,piner14}.
Contrary to this, if \pg\ is assumed to be an intermediate blazar,
the stronger accretion power enabling QPOs scenarios, might be explained by a pulsational perturbational and/or helical structure, also precessional (i.e. geometrical) mechanisms.
Precession can naturally be produced by a binary SMBH companion, with jet wiggling providing a QPO contribution.

On the other hand an intrinsic internal helical structure with a single central SMBH
\citep[e.g., ][]{camezind92,villata99,nakamura04,rieger04} predicts that magnetic fields play a major role in QPOs of blazars \citep[e.g., ][]{mckinney12}. VLBI parsec-scale structures, showing wiggling radio knots \citep[e.g., ][]{piner2010,piner14,lico}, are believed to originate from shocks propagating through cylindrical or conical jets, interacting with a preexisting helical structure. Magnetic field turbulence
can lead to recurring Kelvin–Helmholtz instabilities, capable of generating weak shocks \citep[e.g., ][]{lister13,hughes15}, that are advected into the jet base. The toroidal component of the jet magnetic field is wound up by the accretion flow from the central rotating SMBH. Lense–Thirring (rotational dragging in General Relativity) precession \citep[e.g., ][]{wilkins72,bardeen75,ingram09} of the inner portions of the accretion disk, also can generate QPOs.

The contribution by bulk jet precession, via differential Doppler boosting, is expected to be modest. Changes in direction at the jet nozzle can occur by accretion disk Lense-Thirring precession, or orbital Keplerian motion of the accretion system of a close, $<10^{18}$ cm separation, binary SMBHs.
Parsec-scale jet curvature, wobbling, precession, rotation, nutation (rocking and nodding), or non-ballistic helical motion of, in-jet, emission components are all able to produce flux modulations, via cyclical magnification provided by recurrent Doppler beaming variations. Nozzle and jet base precession and rotation can then produce, particle-accelerating,  MHD stresses and turbulence.
Works following up on A15  reported some of these mechanism for \pg\ \citep{caproni17,sobacchi17,cavaliere17,2019ApJ...875L..22C,tavani2018,sandrinelli18,lico}.

Among many open astrophysical interpretations we list five scenarios for the $\gamma$-ray QPO of \pg:

(a) Pulsational accretion instabilities produce efficiency modulations in the energy outflow. MDAF with subluminal, turbulent and peculiar radio kinematics \citep{fragile09,karouzos12,piner14} could be
explained as a precessing or helical jet \citep{conway93}, and is also able, periodically, to efficiently impart energy to particles in the jet \citep{tchekhovskoy11}. Periodicity here could be too short ($\sim 10^5\,\textrm{s} \cdot M_{SMBH}/10^8\,\textrm{M}_{\odot})$ \citep{honma1992}, but magnetically choked accretion flows can produce longer periods for slow-spinning SMBHs \citep{mckinney12}.

(b) Geometrical jet precession \citep[e.g., ][]{romero00,stirling03,rieger04,caproni13}, rotation and nutation \citep{camezind92,vlahakis98,hardee99,valtonen08} or an intrinsic helical structure where strong winds and a magnetic field wraps the jet \citep[e.g., ][]{conway93,roland94,villata99,nakamura04,ostorero04,raiteri15}, all are able to produce a QPO, through the cyclical change of the viewing angle $\theta$ and change of the bulk Doppler boosting factor $\mathcal{D}=1/\left( \Gamma(1-\beta \cos \theta)\right)$. Here $\beta=v/c$, $\Gamma=1/\sqrt{1-\beta^2}$. The observed flux $F_{\gamma} \propto E^{-\alpha}$ is proportional to $\mathcal{D}^{3-\alpha}$,
for a boosted high-energy emission blob in the jet. In \pg\ a variation $\Delta \mathcal{D}(t) = \Gamma^{-1}(1- \beta(t) \cos \theta(t))^{-1}$ \citep{rieger04} greater than $40\%$ with precession angle $\sim 1\arcdeg$ is required to explain the  $\sim 2.8$ flux modulation factor seen in the $\gamma$-ray light curves obtained by the LAT. On the other hand
$\mathcal{D}$ asympotically diverges for small $\beta \cos \theta$, and $\Delta \mathcal{D}(t)$ changes substantially if the wobbling spans very small angles. In addition, concurrent intrinsic outflow and efficiency variations (point (a)) could help to relax this constraint.

(c) A mechanism analogous to a low-frequency QPO from high-mass X-ray  binaries is supported by the many BH timescales that depend inversely on the mass \citep{fender04,king13}, where the accretion-outflow coupling is the basis of the periodicity. The microquasar mechanism of Lense-Thirring precession of the jet's nozzle \citep{wilkins72} requires that the inner accretion flow forms a geometrically thick and viscous torus. A standard thin disk instead warps \citep[Bardeen-Petterson effect, ][]{bardeen75} rather than precesses \citep{ingram09}. In the case of a binary SMBH system, the primary minidisk precesses with nutation by  the gravitational influence of the secondary, changing jet direction, viewing angle and Doppler boosting factor $\mathcal{D}$. It also changes the variability timescales ($\Delta t / \Delta t' = 1/\mathcal{D}$, but these usually act on much longer time scales than the 2-year period.

(d) Eccentric orbiting massive stars (or an intermediate-mass BH) with inclined/polar revolutions around a single SMBH. This should be very common and could provide periodic modulations of the accretion flow as introduced in Section \ref{sect:bigstar}.

(e) A gravitationally bound binary SMBH system \citep{begelman80,barnes92} with a total mass $5.8 \times 10^8$
$\textrm{M}_{\odot}$ \citep{2021MNRAS.506.1198D}, and a milli-pc separation in the early inspiral gravitational-wave driven regime, as introduced in A15.
Most binary SMBHs will spend the majority of their lifetime at 0.01 – 1 pc separations, in an intermediate phase of evolution between scattering any stars in the nuclear region and gravitational radiation dominance \citep{2015MNRAS.454L..66S}. Keplerian binary orbital motion would induce periodic accretion perturbations \citep{valtonen08,pihajoki13,liu15,2018acps.confE..41C}
or jet nutation  due to the misalignment of the rotating SMBH spins or the gravitational torque on the disk exerted by the companion \citep{katz97,romero00,caproni13,graham15}.

Significant acceleration of the disk evolution and accretion onto a binary SMBH system, where there is excited eccentricity in the inner region of
the circumbinary disk can create an overdense lump giving rise to enhanced periodicity in the accretion rate \citep{2012A&A...545A.127R,nixon13,2014PhRvD..89f4060G,2014ApJ...783..134F,dogan15}.

Binary SMBH-induced periodicities have timescales ranging from $\sim 1$ to $\sim 25$ years  \citep{komossa06,rieger07}.
The  total mass of the SMBH in \pg\ can also be estimated by the putative link between inflow /accretion (disk luminosity) and  outflow (jet power) \citep{ghisellini14} with a 0.1 $dM_{Edd}/dt$ rate and $\mathcal{D}=30$. Comparing luminosities $ L_{disk}$, $L_{rad}$
and jet radiative power $P_{rad}$, we obtain $M_{SMBH} \approx 1.6\times  10^8$ $\textrm{M}_{\odot}$,
in agreement with \citet{woo05}, but a factor of 3.5 lower than that in \citet{2021MNRAS.506.1198D}.
The observed $(2.1\pm0.2)$-year period is
equivalent to an intrinsic orbital time $T'_{Kep} \leq T_{obs}/(1+z) \simeq 1.5 $ years. Considering a Keplerian orbit for the bound binary SMBH gives a binary size of $0.005$ pc $ \simeq 100\textrm{R}_{S}$, with $R_{S}= 2G(M_1)/c^{2}$ the Schwarzschild radius and $M_1$ the mass of the main SMBH. The probability to be observing such a milli-pc system, estimated from the binary mass ratios $\sim 0.1-0.01$ and the GW-driven regime lifetime \citep{peters64}
%
$t_{GW}=(5\ c^5 a^4)/(256\ q\, G^3 \, M^3)
\simeq 10^5 - 10^6 $ years (with $q = M_1 M_2 / (M_1 + M_2)^2 \sim 0.1 - 0.01$)
might be too small. In Section \ref{sect:binarysystem} we briefly evaluate the gravitational lifetime and reliability of the binary SMBH hypothesis, while in Section \ref{GWradiationbackground} we consider the consistency with a binary SMBH population and the gravitational radiation background.
%
%
%

\subsection{Binary SMBH model reliability} \label{sect:binarysystem}

In principle, a binary system of two accreting and jetted SMBHs with
$M_{1}=1.6\times 10^{8} \textrm{M}_{\odot}$ and mass ratio in the range $q=M_{2}/M_{1}=0.01-0.1$, can induce the coherent, high-energy/multifrequency, short ($\sim 2.1$-year) periodicity, with $T'_{Kep}\simeq 1.5 $ year in the source rest frame, as observed in \pg.
The  lifetime of the system, which is directly proportional to its probability of being observed, can be evaluated  as the standard GW merger timescale $t_{GW}$, as defined in the previous section,
where circular orbits are here considered, and separation $a$ depending on the characteristic orbital period $T$ of each of the three hypotheses below.

1) The periodicity of the binary SMBH  causes a periodic accretion supply of gas in the minidisk of the primary SMBH, which is the direct source of the variability of the relativistic jet emission through quadrupolar torque on the large circumbinary disk \citep[e.g., ][]{rieger04,2007ApJ...668..417F,2012A&A...545A.127R,2014ApJ...783..134F,2017MNRAS.469.4258T,2019MNRAS.485.1579K}. This binary Keplerian orbital period $T_{k}$ dictates the observed periodicity and $a=\left( T_{k}^{2}~GM / (4 \pi^{2}) \right)^{1/3}$ yielding the lifetime: $t_{GW} \sim 2 \times 10^{6}$~years for mass ratio $q=M_{2}/M_{1}=0.01$ ($2 \times 10^{5}$~years for $q=0.1$), in agreement with our estimates.

2) Considering the jet aligned with the primary BH spin, the secondary jet precesses around the total angular momentum of the binary spin-orbit (SO) coupling, with the same periodicity. This spin-orbit precession has a$=$
\begin{math}
\left( T_{so}~q~(4+3q)~GM_{1}\sqrt{G(M_{1}+M_{2}}) ~/~(4\pi~(1+q)~c^2) \right)^{2/5}
\end{math}

\citep{1994PhRvD..49.6274A} yielding a too-short lifetime of $t_{GW} \sim 0.8$~years for $q=0.1$ ($0.2$~years for $q=0.01$);  such a system would certainly not persist 15 years.

3) If the primary jet, perpendicular to the SMBH accretion minidisk, is misaligned with the binary orbital plane, the minidisk precesses due to the torque induced by the secondary SMBH with a timescale depending on its distance from the primary SMBH. Torques that tend to align the minidisk with the orbital plane, compete with Bardeen–Petterson torques that tend to align the minidisk with the SMBH spin.
With the jet launched by a magnetic funnel anchored to the inner edge of the disk, the relevant distance is the innermost stable circular orbit (ISCO). With the precession of the primary accretion disk induced by the secondary SMBH, similar to the second case with spin-orbit precession periodicity observed $T_{SO}$, now for the inner ring of the accretion disk, around the ISCO and assumed as the jet launcher, we have the disk semimajor axis: $a_{d}=\left( (3T_{d}~q~(6GM_{1}/c^{2})^{3/2}\sqrt{G(M_{1}+M_{2})}~cos~i)~/\left(8\pi~\sqrt{1+q}\right)\right)^{1/3} $, \citep{2013ApJ...774...43M}, where $i$ is the inclination of the disk at the ISCO (for $i \sim \pi/4$ or smaller it to be yields similar results depending only weakly on $i$), and $T_{d}$ the precession period of the inner ring of the accretion disk. Again, such a system would certainly
not live 15 years ($t_{GW} \sim 0.6$~years for $q=0.1$ and $0.3$~years for $q=0.01$).

\begin{figure}[tttttttt!!]
\centering 
\hskip -0.2cm
\resizebox{\hsize}{!}{\rotatebox[]{0}{\includegraphics{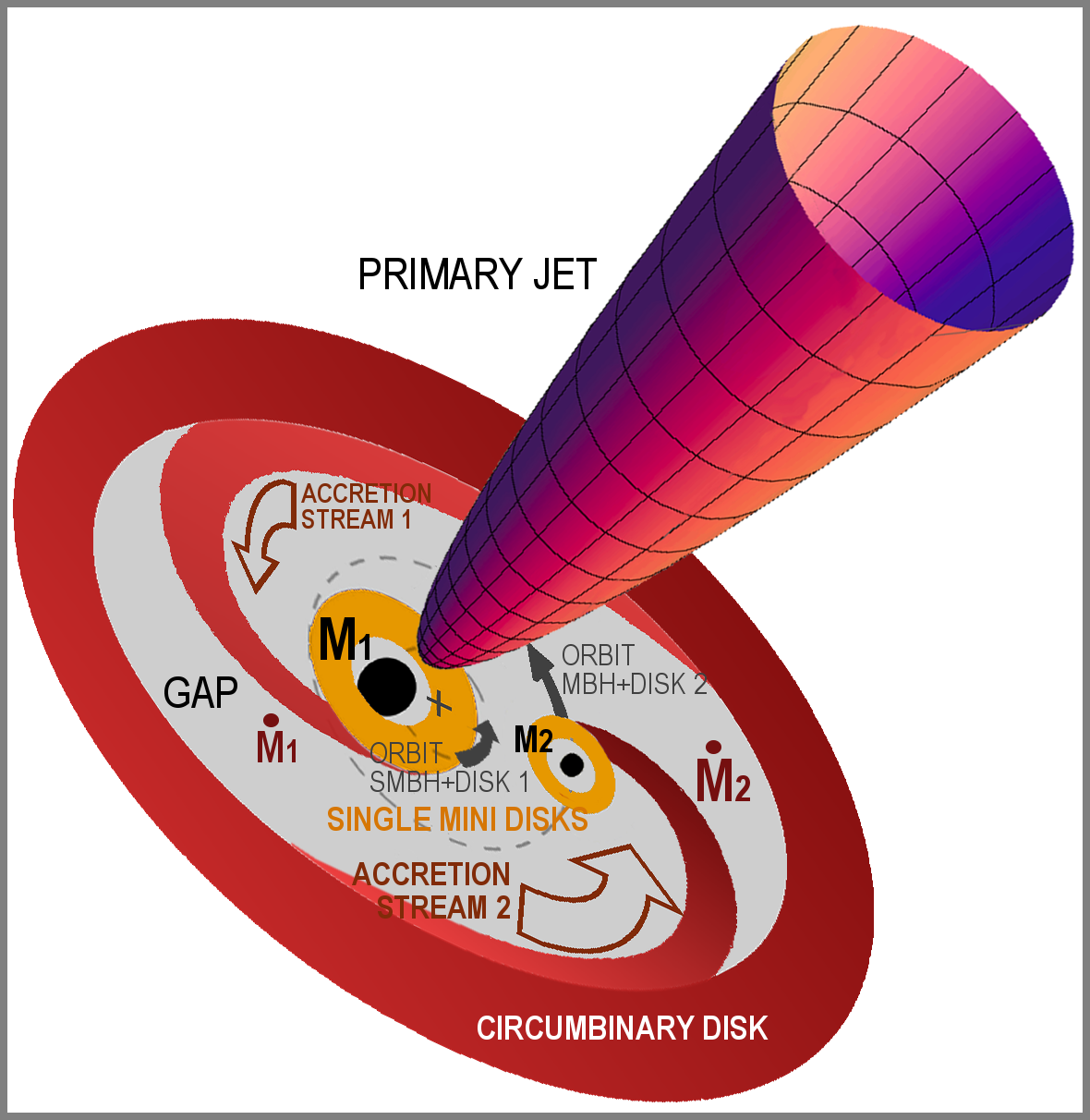}}}
\caption{Cartoon of the hydrodynamical accretion geometries of a binary-SMBH theoretical scenario for \pg. Around each SMBH a ``mini circum-single'' disk is fed by variable accretion streams ($\dot{M_{1}},\dot{M_{2}}$) extending from a large and dense circumbinary disk, having a gap radius of about twice the binary separation. The relatively high mass ratio (e.g. $q=M_{2}/M_{1}\simeq 0.01$ or $q=0.1$) also allows  the secondary, $\#2$ SMBH ($\#2$, $M_{2}\simeq 10^{5}~\textrm{M}_{\odot}$, or  $ 10^{6}~\textrm{M}_{\odot}$ ) to reach the vicinity of the circumbinary annulus, tending to receive a disproportionate share of the accretion rate ($\dot{M_{2}}>\dot{M_{1}}$), causing an additional periodic perturbation. The circumbinary disk also has  intrinsic oscillations by gravitational quadrupolar torque and orbits at longer periods than the two mini-disks feed, producing further cyclic variations in the accretion rate,  jet matter and energy feeding.}
\label{fig:cartoon_binary_model}
\vskip -0.3cm
\end{figure}

The $1.5$-year rest-frame periodicity cannot therefore  be ascribed to precession timescales induced by the
binary companion, either on the primary SMBH spin or on its accretion disk, because of the too-short binary merging timescale ($\lesssim 1$ year).
If we postulate a binary SMBH system in \pg, the only reasonable mechanism is a Keplerian periodic orbital modulation of the accretion rate onto the primary SMBH induced by the companion, which dominates the hydrodynamic periodic variability with a recurrent fuelling of the primary SMBH accretion disk and the activity of its relativistic jet. A tentative graphical toy-model illustration of accretion geometries for such a binary SMBH system is shown in Figure \ref{fig:cartoon_binary_model}.

A periodic perturbation of the primary ($\#1$, $M_{1}\simeq 10^{8}~\textrm{M}_{\odot}$) SMBH's minidisk and accretion power, feeding the primary relativistic jet responsible for $\gamma$-ray emission, is provided through an intrinsically oscillating, gravitational quadrupolar torque of the large and very dense circumbinary disk that couples with the two SMBHs dynamics. The cavity gap of the circumbinary disk has a radius roughly twice the binary separation. Periodic accretion streams ($\dot{M_{1}},\dot{M_{2}}$) extending from the circumbinary disk, orbiting at longer periods than the circum-single minidisks, feed the two BHs. The relatively high mass ratio (e.g., $q=M_{2}/M_{1}\simeq 0.01$ or $q=0.1$) also allows  the secondary,  $\#2$ MBH ($\#2$, $M_{2}\simeq 10^{6}~\textrm{M}_{\odot}$, or  $ 10^{7}~\textrm{M}_{\odot}$ ) to reach the vicinity of the circumbinary annulus, with a consequent larger accretion flow and greater gas received ($\dot{M_{2}}>\dot{M_{1}}$), causing an additional periodic perturbation in accretion power and disk tearing of $\# 1$, during  orbit of $\# 2$  around the center of mass (the $+$ cross in the drawing center). The gravitational lifetime for GW radiation losses of  such a binary system geometry would be $t_{GW} \sim 2 \times 10^{6}$~years for  $q=0.01$, and $2 \times 10^{5}$~years for $q=0.1$.

\hskip -0.2cm

\subsection{Pulsar Timing Array constraints on close binary SMBHs}\label{GWradiationbackground}

It is interesting frame our findings for \pg\ within the recent results presented by the pulsar timing array (PTA) experiments ongoing around the world. In fact, evidence for a common red signal, with properties consistent with a gravitational wave (GW) origin has been found in the latest data releases of the European PTA (EPTA, \citealt{2023A&A...678A..50E}), NANOGrav \citep{2023ApJ...951L...8A}, the Parkes PTA (PPTA, \citealt{2023ApJ...951L...6R}) and the Chinese PTA (CPTA, \citealt{2023RAA....23g5024X}). The significance of the detection is between 2$\sigma$ and 4$\sigma$ depending on the dataset, and the amplitudes of the signals are consistent with a stochastic GW background (GWB) produced by a cosmic population of SMBH binary systems (SMBHB), although an early-Universe origin cannot be excluded \citep{CGW,2023ApJ...951L..11A}. Assuming a SMBHB origin of the signal, \citet{2018MNRAS.481L..74H} constructed a theoretical framework to translate its amplitude into the fraction of expected periodic sources observed by \textit{Fermi}-LAT. The details of the model can be found in their paper, but the main result is shown in their Figure 4. This figure shows that, under the assumption of a GWB signal amplitude of $A_{1yr}=1.45\times10^{-15}$
(the published limit at the time), the expected fraction of \fermilat BL Lacs showing flux oscillations with a period less than 5 years is about $3\times 10^{-4}$. The signal now observed in the PTA data has an amplitude of about $A_{1yr}=2.5\times10^{-15}$. Since the amplitude scales with the square root of the number of SMBHBs contributing to the signal, this implies an expected fraction of periodic BL Lacs of about $10^{-3}$.

This is consistent with finding one confirmed periodic source (e.g., \pg ) out of thousands  sources in the \fermilat catalog. It should be noted, however, that \pg\ is among the brightest sources in the \fermilat sample. Moreover, periodicity (albeit at a lower significance) has been suggested for PKS 2155$-$304 and BL Lac \citep{2018A&A...615A.118S} and for a few other systems \citep{2017MNRAS.471.3036P,2020ApJ...896..134P,
2024MNRAS.52710168P,2024MNRAS.529.1365P}, all of which are in the bright flux tail of the \fermi source distribution  \citep[see Figure 4 in][]{2018A&A...615A.118S}. So it might be that the detection of periodicity in \fermilat data is flux limited, and the identification of \pg\ (plus perhaps few other systems) as the only periodic \fermilat sources, is  due to selection effects. Therefore, no firm indication either in favor or against \pg\ being a binary SMBH system can be drawn from the GW signal observed in the PTA data.

\subsection{Perturbation by inclined orbit  star-like objects} \label{sect:bigstar}
Eccentric orbiting star-like objects, or intermediate-mass BH around a single SMBH should be common. With an orbital binding energy $-3GM^{2}/(5R) \simeq $~some MeV per nucleon (stellar interiors kinetic energies of nuclei have a Maxwellian internal energy distribution with $\sim $~keV averaged value), the  Roche density of the star $\rho_R=3.5M/r^{3}$ required to resist tidal heating and disruption is very small for the intrinsic $\gamma$-ray 1.5-year periodicity of \pg\ in the case of non-polar orbits. Highly-inclined stellar orbits of $\sim~10^{16}$ cm radius can be perturbed more easily than a MBH/SMBH orbit, with a gravitational energy loss and orbit co-alignment with the disk timescale of $>10^{9}$ years \citep[$10^3\times$ the stellar interaction time; for example ][]{2013MNRAS.434.2948D,2023ApJ...957...34L}.

Local fire bubbles ($\sim 10^{5}$~K thermal bremsstrahlung) or magnetic reconnection flare avalanches in the disk and corona are expected from disk piercing, twice, by the star (or WD, NS or star-sized BH) with energy release $\Delta E = \frac{1}{2}(\Sigma \sigma^2 v_r^2)$; where $\Sigma$ is the disk surface density, $\sigma^2$ the disk-star cross section and $v_r$ the hypersonic relative velocity of the star to the disk matter. The disk serves not only as the source of the outflow material, but it also illuminates it and reflects the outflow radiation back.
Accretion flow perturbations propagating to the inner disk are the result of double disk-star impact fireballs, modulating the accretion rate, nozzle and wind-confining relativistic jet of a blazar like \pg\,  \citep{1998ApJ...507..131I}. General Relativity MHD simulations, with an orbiting intermediate BH (IMBH, $10^2-10^5$ M$_\odot$) companion, indicates mildly relativistic particle acceleration in the stable poloidal magnetic field of the jet funnel, and ADAF ejecting outflow clumps  \citep{pasham2024,2024arXiv240400941K}. High eccentricity of the IMBH's orbit around the SMBH, can also explain variations from a strict periodicity.

\section{Conclusions} \label{sect:conclusions}

The primary aim of this paper is to follow up the tentative identification of a $2.1\pm0.2$ year periodic oscillation in the blazar \pg. The additional data used in this study provide further and independent support for the existence of the periodic flux modulation, which was described in the initial study in A15.

The Fermi-LAT $\gamma$-ray data show a persistent 2.1-year oscillation throughout the 15 years of observations. Compared to the A15 analysis, the QPO significance has increased to $4 \sigma$ with respect to simulations using two different types of red noise models.

An essential benefit of the longer time series is that it allows us to compare oscillation properties in two independent parts of the data. Both sections of the light curve are found to contain an oscillation with well-determined period and phase. The chance probability for the two oscillations to be so close in period and phase is less than 0.01.

A multi-wavelength analysis is used to study correlations between the $\gamma$-ray flux and the X-ray, optical and radio measurements. No significant $\gamma$-ray flux correlation is found with X-rays, while both optical and OVRO 15 GHz radio show strong correlations with time lags $6 \pm 18$ and $188 \pm 28$ days, respectively, relative to the \textit{Fermi}-LAT ($E>1$ GeV) $\gamma$-rays. Pulse timing over 8  $\gamma$-ray periodic cycles shows arrival time variations that are consistent with a strictly coherent period. Additional years of long-term, and regular, flux monitor of \pg\ are needed, especially at X-ray and VHE $\gamma$-ray ($E>100$ GeV) energy bands, possibly with a regular radio/optical polarization monitor.

A spectral $\gamma$-ray analysis reveals an energy dependence of the QPO such that its amplitude is greater above 1 GeV than at lower energies. We also find hints for softening during a precursor feature in the rising part of the oscillation pulse.

The binary SMBH interpretation, with $1.6 \times 10^{8}$ M$_{\odot}$ total mass, $\sim 100~\mathrm{R}_{s}$  separation and lifetime (residence time $\sim |a/(da/dt)| \sim 10^{5},10^{6}$ years, where $a$ is the semi-major axis of the binary) has been discussed
in Section 4, although at least four additional astrophysical scenarios, including precessing, nutation or helical and wiggling jet structures, MHD instabilities and perturbations by inclined massive stellar objects or intermediate BHs,  are possible.
The $\sim 1.5$-year rest-frame QPO of \pg\ cannot be ascribed to precession timescales (binary spin-orbit coupling  or spin/accretion disk precession of the primary SMBH induced by the lower-mass  companion), because of the too-short ($\lesssim 1$ year) merging scale with these mechanisms. The optical-$\gamma$-ray QPO can be caused by Keplerian periodic orbital modulation of the supplied gas and matter in the main jet by the primary minidisk of the SMBH, with the contribution also of a substantial quadrupolar torque of the large and dense circumbinary disk.
The QPO might be a hydrodynamic periodic variability of the fuelling of the primary SMBH accretion power, where the toroidal component of the jet magnetic field is wound up by the accretion flow, and magneto-rotational stresses are induced in the jet, energizing electron populations and producing an underlying periodic modulation in $\gamma$-ray flux enhancements. This would survive for the rather long gravitational-losses lifetime, prior to the disruptive coalescence of
$t_{GW} \sim 2 \times 10^{6}$~years for  $q=0.01$, and $2 \times 10^{5}$~years for $q=0.1$. The case of \pg\  (chirp mass  ${\cal M}_8\lesssim1$), representing the strongest case of significant QPO out of a sample of $\sim1000$ \textit{Fermi}-LAT bright $\gamma$-ray BL Lac objects, could suggest  possible contribution to the GW stochastic background probed by international PTAs, from the entire object class of BL Lacs and their misaligned AGN parent population. Instabilities and perturbations are easier in the radiatively inefficient regimes of BL Lac objects, and their supposed longer lifetimes $t_{GW}$ in close orbits, with
respect to those of heavier and more distant quasars and FSRQs. This also supports a GW stochastic contribution, which could be determined in the future by SKA. There is already a candidate detection of 4-5 nHz continuous GW emission generated by  binary SMBHs in the local Universe \citep{CGW}.

Minor mergers, i.e. $M_2\lesssim M_1/4$, are more likely to be observed electromagnetically, because the time to merge is $t_{\rm GR}\propto M_2^{-1}M_1^{-2/3}$ and therefore these should be more frequent events. In this view, perturbations by a highly eccentric and inclined polar-orbiting massive star or an intermediate-mass BH are interesting to consider for the case of \pg.

QPO and related astrophysical models for \pg\ could be evaluated in the future by observations of helical or wiggling and wobbling patterns in the pc-scale radio jet, by the identification of regular radio-components and polarization patterns and by double-peaked spectroscopic emission lines. The recent X-ray observations of \pg\ by {\it IXPE} and simultaneous multifrequency data in a range of $\sim8$ days in February 2023 \citep{middei23}
point out turbulence as a contributor to the jet magnetic field that, eventually, may produce  instabilities generating weak shocks advected into the jet base. {\it TESS} optical observations over a range of 24 days in April-May 2022 could add more information about jet turbulence.

The unique \fermilat\ all-sky and time-domain survey will continue, providing essential data for \pg\ and triggering the scientific interest in this potential multimessenger science case, while also promoting efforts in multifrequency follow-up observations.



\acknowledgments
$~$

We thank the anonymous referee for useful comments.
The \textit{Fermi}-LAT Collaboration acknowledges generous ongoing support from a number of agencies and institutes that have supported both the development and the operation of the LAT as well as scientific data analysis. These include the National Aeronautics and Space Administration (NASA) and the Department of Energy (DoE) in the United States, the Commissariat \`a l'Energie Atomique (CEA) and the Centre National de la Recherche Scientifique / Institut National de Physique Nucl\'eaire et de Physique des Particules (CNRS IN2P3) in France, the Agenzia Spaziale Italiana (ASI) and the Istituto Nazionale di Fisica Nucleare (INFN) in Italy, the Ministry of Education, Culture, Sports, Science and Technology (MEXT), High Energy Accelerator Research Organization (KEK) and Japan Aerospace Exploration Agency (JAXA) in Japan, and the K.~A.~Wallenberg Foundation, the Swedish Research Council (Vetenskapsr{\aa}det,  VR) and the Swedish National Space Board (SNSB) in Sweden. Additional support for science analysis during the operations phase is gratefully
acknowledged from the Istituto Nazionale di Astrofisica (INAF) in Italy and the Centre
National d'\'Etudes Spatiales (CNES) in France. This work performed in part under DOE Contract DE-AC02-76SF00515.

The NASA satellite \textit{Neil Gehrels Swift} $\gamma$-ray burst explorer is a MIDEX Gamma Ray Burst mission led by NASA with participation of Italy and the UK. The NASA satellite \textit{Rossi} X-ray Timing Explorer (RXTE) was a mission  managed and controlled by NASA's Goddard Space Flight Center (GSFC) in Greenbelt, Maryland. The RXTE all-sky monitor (ASM) light curves project was possible thanks to RXTE teams at MIT and GSFC, and supported by NASA contract No.NAS5-30612.

This research has made use of data from the OVRO 40-m monitoring program \citep{2011ApJS..194...29R} which is supported in part by NASA grants NNX08AW31G, NNX11A043G, and NNX14AQ89G and NSF grants AST-0808050 and AST-1109911.
The MOJAVE program is supported under NASA-FERMI grant NNX12A087G.
The National Radio Astronomy Observatory (NRAO) is a facility of the National Science Foundation operated under cooperative agreement by Associated Universities, Inc.
This work has made use of data obtained by the Tuorla optical blazar monitoring program (mainly KVA observatory and the 1.03-meter telescope at Tuorla Observatory), in this case all included in paper A15. This work has made use of data obtained with the 0.76-meter Katzman Automatic Imaging Telescope (KAIT), supported by donations from corporations, foundations and the Lick Observatory and the U.S. NSF. This work has made use of data obtained with the Catalina Real-Time Transient Survey (CRTS), supported by the U.S. NSF under grants AST-0909182 and AST-1313422, and the CSS survey funded by the NASA Grant No.NNG05GF22G. This work has made use of data obtained with ASAS-SN project, supported by a foundation grant GBMF5490 to the Ohio State University and NSF grants AST-1515927 and AST-1908570. The DASCH project at Harvard is grateful for partial
support from U.S. NSF grants AST-0407380, AST-0909073, and AST-1313370.

This research has made use of data and/or software provided by the High Energy Astrophysics Science Archive Research Center (HEASARC), which is a service of the Astrophysics Science Division at NASA/GSFC. This research has made use of  the Smithsonian/NASA's ADS bibliographic database. This research has made use of the NASA/IPAC Extra-
galactic Database (NED) database which is operated by the Jet Propulsion Laboratory, California Institute of Technology, under contract with NASA. This research has made use of archival data, software, and/or
online services provided by the Space Science Data Center (SSDC) operated by ASI.
This research has made use of the XRT Data Analysis Software (XRTDAS) developed under the responsibility of the SSDC. This research has made use of the SIMBAD database, CDS, Strasbourg Astronomical Observatory, France.

This work was supported by a grant from the Simons Foundation (00001470, R.B.).
S.Ci. and INFN personnel in SSDC also performed under contract agreement ASI-INFN 2021-43-HH.0.
S.Ge acknowledge Italian MUR, program “Dipartimenti di Eccellenza 2018-2022” (Grant SUPER-C).

Facilities: \textit{Fermi Gamma-ray Space Telescope} --- \textit{LAT}  --- \textit{Swift}  --- \textit{XRT} --- \textit{UVOT} ---
\textit{RXTE} ---
\textit{OVRO}  --- \textit{Tuorla} --- \textit{KVA} --- \textit{KAIT} --- \textit{CSS} --- \textit{CRTS} --- \textit{ASAS-SN}

\normalsize

\bibliographystyle{apj}

{}


\appendix
\section{The 3-day and weekly bin public LAT light curves}\label{sect:3daybinlightcurveanalysis}

\begin{figure*}[ttthhhhhhtt!!]
\hskip -0.8cm
\resizebox{1.1\hsize}{!}{\rotatebox[]{0}{\includegraphics{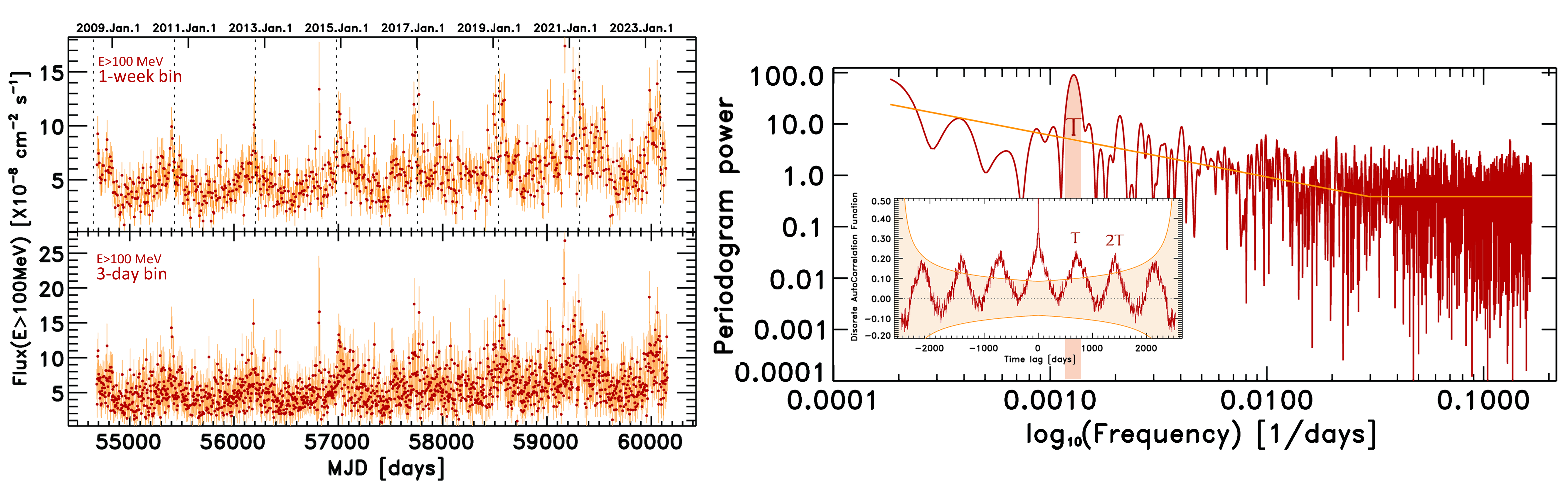}}}
\caption{Left panel: 1-week bin and 3-day bin, science-ready and public $\textit{Fermi}$-LAT light curve of \pg~ from 2008 August 8 to 2023 November 16, i.e. from 15 years of \fermi  mission survey operations, extracted from the LAT light curves repository (LCR) at the \fermi Science Support Center (GSFC-NASA). Right panel:  Corresponding periodogram power density spectrum (main panel) from the 3-day bin light curve , and DACF in inset, with the periodic component $T=2.13$ years highlighted. This period is consistent with the findings reported in the main text of this paper.}
\label{fig:DACF45d_DACFSFLSP3d}
\vskip 0.3cm
\end{figure*}

\begin{figure}[bbbbbhhhhhhbb!!]
\centering 
\hskip -0.8cm
\resizebox{0.45\hsize}{!}{\rotatebox[]{0}{\includegraphics{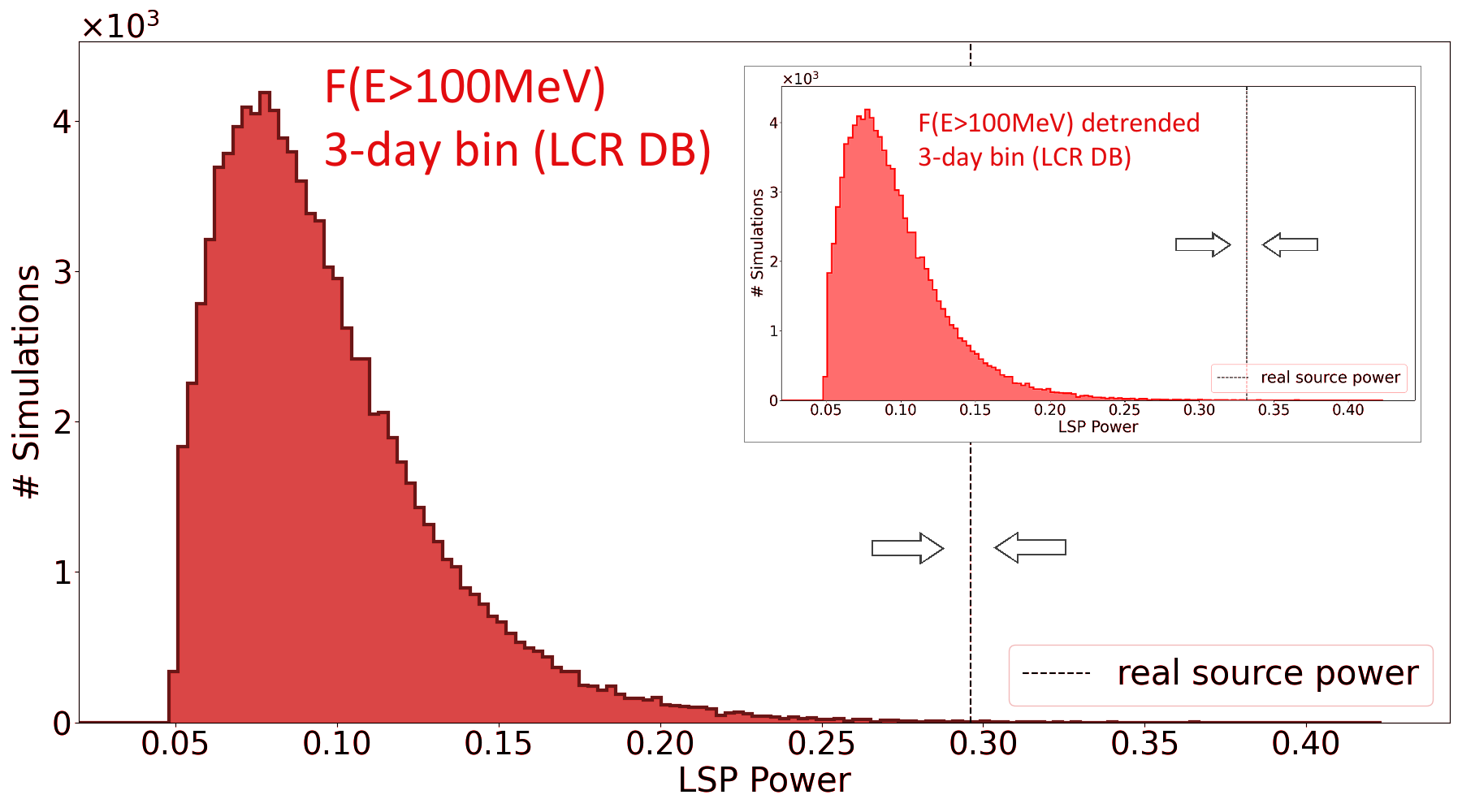}}}
\vskip -0.2cm
\caption{Distribution of the LSP signal peaks for $10^6$ simulated light curves, respectively for the public 3-day bin LCR light curve shown in Figure \ref{fig:DACF45d_DACFSFLSP3d} in the main panel, and for the detrended version of it in the inset panel. Dashed lines are the
LSP power for the two, true data, light curves. Significance of the
period here is $4\sigma$ and $>4\sigma$ for the 3-day bin LCR light curve and the detrended version of it, respectively.}
\label{fig:3daybinLCR_emmansimulDistrib}
\vskip 0.3cm
\end{figure}

The evidence of a potential  $\sim 2.1$-year   periodic $\gamma$-ray flux modulation is independent of the bin value adopted to calculate the \textit{Fermi}-LAT light curve of \pg\. As an example here the analysis of the public \textit{Fermi} LAT light curve data of \pg\, from  2008 August to 2023 November, with a fine, 3-day binning and publicly available at the
\fermilat
light curve repository of the
FSSC\footnote{\texttt{https://fermi.gsfc.nasa.gov/ssc/data/access/lat/LightCurveRepository/\\source.html?source\_name=4FGL\_J1555.7+1111}}, \citep{kocevski21,LCR})
is shown in Figure \ref{fig:DACF45d_DACFSFLSP3d}.
In addition, other techniques are able to highlight  such a signal of periodic oscillation, including the phase dispersion minimization, the discrete autocorrelation function (DACF), and the structure function. The LSP spectrogram and DACF are included in the right panel of  Figure \ref{fig:DACF45d_DACFSFLSP3d}. The LSP power spectrum shows a break and flattening at about $10\times$ the light curve bin size (34 days), and a signal power-law index $\alpha = 0.8$ ($1/f^{\alpha}$ spectrum, $f=1/t$) for longer timescales. The power peak at $2.13 \pm 0.22$ years is also evident. $T$ in the main plot and inset (DACF) represent the $\sim 2.1$-year timescale.

\newpage

\section{1996-2011 RXTE ASM light curve and 1912-1988 historical optical light curve}\label{sect:appendixopticalhistorical}

\begin{figure*}[tttttthhhhhh!!!] 
\centering 
\hskip -0.4cm
\resizebox{\hsize}{!}{\rotatebox[]{0}{\includegraphics{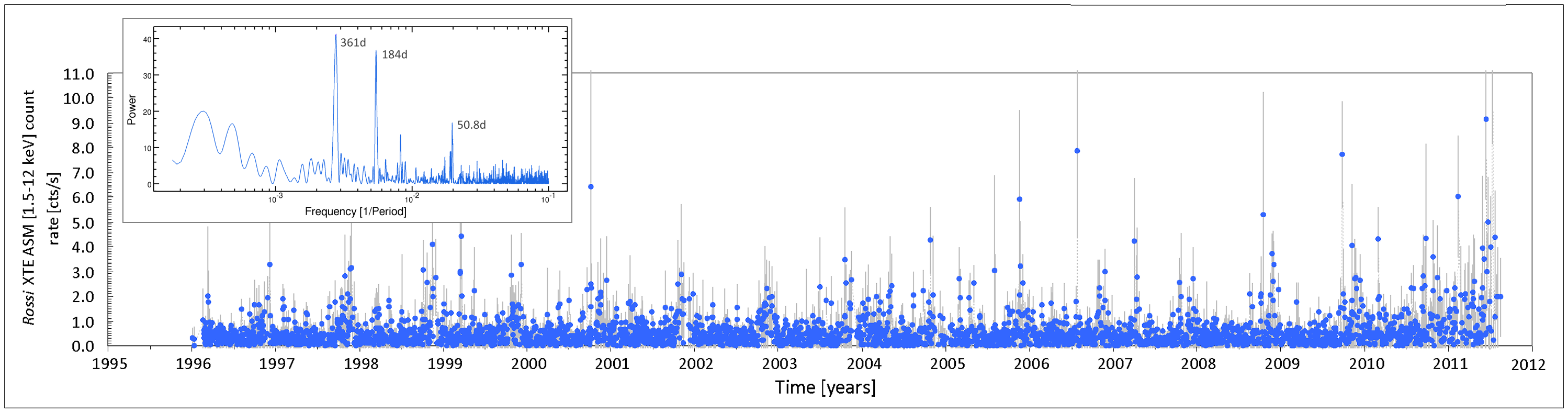}}}
\\[0.1cm]
\hskip -0.4cm
\resizebox{\hsize}{!}{\rotatebox[]{0}{\includegraphics{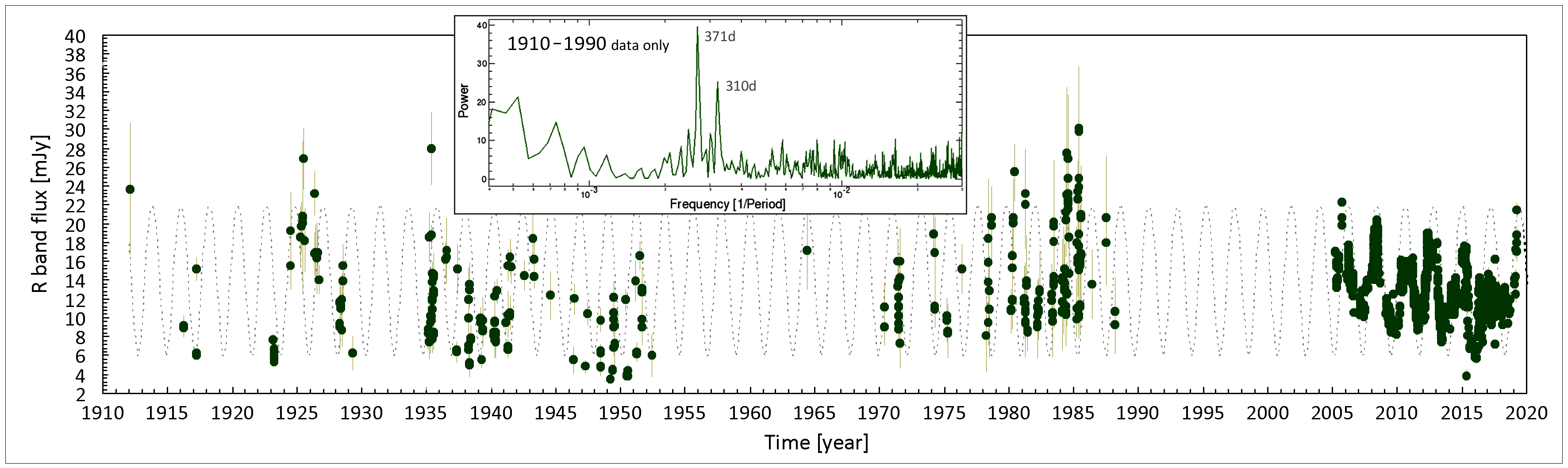}}}
\vskip -0.2cm
\caption{Top panel: the \textit{Rossi} XTE ASM 1.5-12 keV X-ray count rate light curve with 1-day bin averaged data points from 1996 to 2011 (a range of 15.7 years) with negative rates and upper limits removed. In the inset panel the corresponding Lomb-Scargle periodogram showing only spurious 1-year periodicity owing to data gaps due to the Sun vicinity with \pg, its 1/2-year harmonic and the 50-day scale of the typical Sun gap duration and the spacecraft precession.
Bottom panel: Secular historical optical flux light curve of \pg\ with data points from 1912 to 2020 obtained using DASCH DR5 with all data scaled with constant color indexes to approximately the  $R$-band flux densities. A strict 2.1-year sinusoidal line is added as background. In the inset panel the Lomb-Scargle periodogram for the 1910-1990 historical and highly-gapped portion of the light curve is shown, highlighting only low-frequency noise and the usual 1-year timescale (mild peaks at 371 and 310 days), common in ground-based optical light curves. The data after the 2005 were already presented in Sections \ref{optical-data} and \ref{xray-data}.}
\label{fig:historical}
\vskip -0.2cm
\end{figure*}

1-day bin averaged count rate X-ray (1.5-12 keV) light curve calculated from the All-Sky Monitor (ASM) archive\footnote{https://heasarc.gsfc.nasa.gov/docs/xte/} \citep{1996ApJ...469L..33L} of the {\it Rossi X-Ray Timing Explorer satellite (RXTE)}, from  1996 January 6, to  2011 September 23 (Figure \ref{fig:historical}). Only spurious signal power peaks ( $\sim 1$
year, $\sim 1/2$ year, and $\sim 50$ days) are found.

\pg\ is an optically bright blazar (range $ 15\lesssim R \lesssim 13$ mag), with non-negligible chances for detection in optical observatory plates, despite the fact that it is not very close to relevant, historically intensively observed, M/NGC/IC catalog objects in the field. In Figure \ref{fig:historical}, we present a first version of the \pg\ secular optical light curve (1912-1988) extracted from the ``Digital Access to a Sky Century at Harvard'' (DASCH) project for  plate scanning and digitization, and based on 450000 plates (years from 1890 to 1990) of the Harvard College Observatory \citep{2012IAUS..285...29G}. $R$-band flux densities (1912 to 1988) from DASH DR 5 and magnitudes calibrated with APASS AAVSO Photometry All-Sky Survey Rel. 8 ($B$-band mag) and ATLAS All-Sky Stellar Reference Catalog Ver. 2 ($G \simeq V$ band mag), are extrapolated using fixed color indexes $B-R=0.55$ and $B-V=0.25$. The various periodicity analysis methods are not successful.
Future searches for plates and scanning projects around the world would likely help to improve this large-gapped and sparse historical timeseries.

\end{document}

%% file: authorsOKfinal_4july2024.tex
\author[0000-0002-6803-3605]{S.~Abdollahi}
\affiliation{IRAP, Universit\'e de Toulouse, CNRS, UPS, CNES, F-31028 Toulouse, France}
\author[0000-0002-9785-7726]{L.~Baldini}
\affiliation{Universit\`a di Pisa and Istituto Nazionale di Fisica Nucleare, Sezione di Pisa I-56127 Pisa, Italy}
\author{G.~Barbiellini}
\affiliation{Istituto Nazionale di Fisica Nucleare, Sezione di Trieste, I-34127 Trieste, Italy}
\affiliation{Dipartimento di Fisica, Universit\`a di Trieste, I-34127 Trieste, Italy}
\author[0000-0002-2469-7063]{R.~Bellazzini}
\affiliation{Istituto Nazionale di Fisica Nucleare, Sezione di Pisa, I-56127 Pisa, Italy}
\author[0000-0002-4551-772X]{B.~Berenji}
\affiliation{California State University, Los Angeles, Department of Physics and Astronomy, Los Angeles, CA 90032, USA}
\author[0000-0001-9935-8106]{E.~Bissaldi}
\affiliation{Dipartimento di Fisica ``M. Merlin" dell'Universit\`a e del Politecnico di Bari, via Amendola 173, I-70126 Bari, Italy}
\affiliation{Istituto Nazionale di Fisica Nucleare, Sezione di Bari, I-70126 Bari, Italy}
\author[0000-0002-1854-5506]{R.~D.~Blandford}
\affiliation{W. W. Hansen Experimental Physics Laboratory, Kavli Institute for Particle Astrophysics and Cosmology, Department of Physics and SLAC National Accelerator Laboratory, Stanford University, Stanford, CA 94305, USA}
\author[0000-0002-4264-1215]{R.~Bonino}
\affiliation{Istituto Nazionale di Fisica Nucleare, Sezione di Torino, I-10125 Torino, Italy}
\affiliation{Dipartimento di Fisica, Universit\`a degli Studi di Torino, I-10125 Torino, Italy}
\author[0000-0002-9032-7941]{P.~Bruel}
\affiliation{Laboratoire Leprince-Ringuet, CNRS/IN2P3, \'Ecole polytechnique, Institut Polytechnique de Paris, 91120 Palaiseau, France}
\author[0000-0002-3308-324X]{S.~Buson}
\affiliation{Institut f\"ur Theoretische Physik and Astrophysik, Universit\"at W\"urzburg, D-97074 W\"urzburg, Germany}
% ...
